\documentclass[
a4paper,11pt,
amsmath,
amssymb,accepted=2024-09-02]{quantumarticle}
\pdfoutput=1

\usepackage[numbers]{natbib}
\usepackage{graphicx}
\usepackage{dcolumn}

\usepackage{soul}
\sethlcolor{pink}
\usepackage[caption=false]{subfig}
\usepackage{floatrow}
\usepackage[dvipsnames]{xcolor}
\usepackage{tikz}
\usetikzlibrary{arrows}
\graphicspath{ {./figures/} }
\usepackage{physics}

\usepackage{subfig}

\usepackage{url}
\usepackage{hyperref}
\hypersetup{
    colorlinks=true,
    linkcolor=blue,
    citecolor=blue, 
    filecolor=magenta,      
    urlcolor=blue,
}

\usepackage{mleftright} 

\newcommand{\figref}[1]{\mbox{Fig.~\ref{#1}}}

\renewcommand{\eqref}[1]{\mbox{Eq.~(\ref{#1})}}

\newcommand{\figpanel}[2]{Fig.~\hyperref[#1]{\ref*{#1}(#2)}}
\newcommand{\figpanels}[3]{Fig.~\hyperref[#1]{\ref*{#1}(#2)-(#3)}}
\newcommand{\figpanelNoPrefix}[2]{\hyperref[#1]{\ref*{#1}(#2)}}

\usepackage{units}

\begin{document}

\title{Exact results on finite size corrections for surface codes tailored to biased noise}

\author{Yinzi Xiao}
\affiliation{Department of Computer Science, Paderborn University, Paderborn, Germany}
\email{yinzi.xiao@uni-paderborn.de}
\author{Basudha Srivastava}
\affiliation{Department of Physics, University of Gothenburg, Gothenburg, Sweden}
\affiliation{Quantinuum, Terrington House, 13-15 Hills Rd, Cambridge, CB2 1NL, UK}
\email{basudha.srivastava@quantinuum.com}
\author{Mats Granath}
\affiliation{Department of Physics, University of Gothenburg, Gothenburg, Sweden}
\email{mats.granath@physics.gu.se}
\date{}

\begin{abstract} 
The code-capacity threshold of a scalable quantum error correcting stabilizer code can be expressed as a thermodynamic phase transition of a corresponding random-bond Ising model. Here we study the XY and XZZX surface codes under phase-biased noise, $p_x=p_y=p_z/(2\eta)$, with $\eta\geq 1/2$, and total error rate $p=p_x+p_y+p_z$. By appropriately formulating the boundary conditions, in the rotated code geometry, we find exact solutions at a special disordered point, $p=\frac{1+\eta^{-1}}{2+\eta^{-1}}\gtrsim 0.5$, for arbitrary odd code distance $d$, where the codes reduce to one-dimensional Ising models. The total logical failure rate is given by $P_{f}=\frac{3}{4}-\frac{1}{4}e^{-2d_Z\,\text{artanh}(1/2\eta)}$, where $d_{Z}=d^2$ and $d$ for the two codes respectively, is the effective code distance for pure phase-flip noise. As a consequence, for  code distances $d\ll \eta$, and error rates near the threshold, the XZZX code is effectively equivalent to the phase-flip correcting repetition code over $d$ qubits. 
The large finite size corrections for $d_Z<\eta$ also make threshold extractions, from numerical calculations at moderate code distances, unreliable. We show that calculating thresholds based not only on the total logical failure rate, but also independently on the phase- and bit-flip logical failure rates, gives a more confident estimate. Using this method for the XZZX code with a tensor-network based decoder and code distances up to $d\approx 100$, we find that the thresholds converge to a single value at moderate bias ($\eta=30, 100$), at an error rate above the hashing bound.  

\end{abstract}

\maketitle

\section{\label{sec:introduction}Introduction}
With the recent advances in hardware implementations of quantum error correcting codes~\cite{Kelly2015StateCircuit,Takita2017ExperimentalQubits,PhysRevA.97.052313, wootton2020benchmarking,Andersen2020RepeatedCode, Satzinger2021Realizing, Egan2021Fault-tolerantQubit, Chen2021ExponentialCorrection, Erhard2021EntanglingSurgery, ryananderson2021realization, marques2021logicalqubit, Postler_2022, krinner2022realizing,Bluvstein2021,google2023suppressing,moses2023race,sundaresan2023demonstrating, Gupta2023Encoding, Brown2023Advances, Wang2023FT, Bluvstein2023}, important steps have been taken towards fault-tolerant quantum computing~\cite{PhysRevA.32.3266,PhysRevA.52.R2493,RevModPhys.87.307,Girvin_2023}. Most experiments to date focus on topological codes, such as the surface code~\cite{Kitaev2003Fault-tolerantAnyons,dennis2002topological,PhysRevLett.98.190504,PhysRevA.86.032324}, where stabilizers, in the form of local parity checks, can be laid-out on a two-dimensional grid of qubits. The surface code, in the qubit-efficient ``rotated'' implementation~\cite{Bombin2007OptimalStudy}, encodes a single logical qubit in the Hilbert space of $d\times d$ physical qubits, such that a logical operation requires at least $d$ single qubit operations, giving the code distance of the code. The logical failure rate, corresponding to logical bit- and phase-flip errors, is exponentially suppressed with code distance, provided error rates are below an implementation-specific threshold, thus providing a scalable pathway to very low levels of noise. 

In parallel with experimental developments, new error correcting codes that cater to hardware specificities such as low qubit connectivity~\cite{Chamberland2020, Hastings2021DGLT, Wootton2015, Wootton2021}, or biased single-qubit noise profiles~\cite{PhysRevLett.124.130501,PhysRevX.9.041031,Ataides2021XZZX, Srivastava2022xyzhexagonal, PRXQuantum.2.030345,tiurev2023correcting, Huang2022, Tiurev2023Domain}, as well as codes with a higher density of logical qubits per physical qubit~\cite{gottesman2014faulttolerant,PRXQuantum.2.040101,bravyi2023highthreshold}, are being developed and explored theoretically. A primary feature that characterizes scalable error-correcting codes is the code-capacity threshold, which assumes optimal conditions, including noise-free measurements and independent and identically distributed (iid) single-qubit noise. The code-capacity threshold quantifies the highest possible error tolerance of the code, providing an implementation-agnostic fingerprint. 
Recently it was realized that the surface code can be tailored to phase-biased noise by means of  simple modifications of the stabilizers, resulting in a dramatic increase of the threshold~\cite{Tuckett2018UltrahighNoise, Ataides2021XZZX, dua2022clifford}. In fact, evidence suggests that certain tailored codes, such as the XY and XZZX codes, may have thresholds that equal or surpass the hashing bound, which sets a lower bound on the threshold of a random stabilizer code under any iid Pauli channel~ \cite{PhysRevA.54.3824,wilde2011classical}. (See, Appendix \ref{app:hashing}.) The fact that the standard surface code has significantly lower threshold for phase-biased noise, makes the discovery of non-random codes with such high thresholds encouraging and potentially important for future developments.  

To numerically identify the threshold requires an accurate decoder that maps a syndrome, i.e.\ a set of violated stabilizers, to the most likely logical coset of errors consistent with the syndrome. In addition, and in close analogy to thermodynamic phase transitions, the exact threshold is only manifested in the thermodynamic limit, $d\rightarrow\infty$~\cite{dennis2002topological,wang2003confinement}. In this work we quantify the finite size corrections for the XY and the XZZX codes, partly based on exact solutions for the logical failure rate at isolated points in the noise parameter-space, and show that to approach the thermodynamic limit in simulations may require exceedingly large code distances. 
Thus, even with an accurate ``maximum-likelihood decoder''~\cite{Wootton2012HighCode, Hutter2014EfficientCode, Bravyi2014EfficientCode}, it may be difficult to extract correct thresholds, given the large finite-size corrections. In particular, for highly biased noise, the standard method of fitting the logical failure rate using a scaling form cannot be trusted. We show that extracting separate thresholds for logical bit- and phase-flip errors, and study their convergence with respect to code distance, gives a more confident threshold estimate. This methodology can be used for any code that is known to have a unique threshold. For the XZZX model, the analysis shows, quite conclusively, that the code-capacity threshold for moderately biased noise is above the hashing bound.  

The paper is organized as follows. In Section \ref{sec:background} we introduce the XY and XZZX models and the phase-biased single qubit error channel. We review decoding and the Pauli channel for logical errors. In Section \ref{sec:mapping} we review the mapping to generalized Ising models, and specify the boundary conditions for the rotated code geometries. Subsequently, in Section \ref{sec:exact} we derive exact solutions to the logical failure rates of the two models at a special disordered point, and compare this to numerical simulations. In Section \ref{sec:results} we review the finite size scaling fits of thresholds and present numerical results for the convergence of the logical failure rates with code-distance. In Section \ref{sec:low_weight} we make an explicit construction of logical bit-flip operators with a single minority error, which give the leading contribution to the bit-flip error rate for large bias and high error rates. In Section \ref{sec:discussion} we discuss the implications of the results in light of earlier studies of surface codes tailored to highly biased noise, and in Section \ref{sec:conclusion} we conclude. Appendices give a review of the hashing bound for the code-capacity threshold of a random stabilizer code (\ref{app:hashing}), a review of the 1D Ising model (\ref{app:one-d-ising}), an excursion to the standard (XZ) surface code under biased noise (\ref{app:xz_surface}), and  present extended results and details of the numerical data and threshold fits (\ref{app:thresholds}).     

\section{\label{sec:background}Background}

\begin{figure}\centering

\subfloat[]{\includegraphics[width=.5\linewidth]{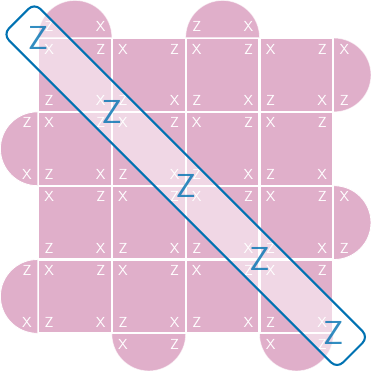}}\hfill
\subfloat[]{\includegraphics[width=.5\linewidth]{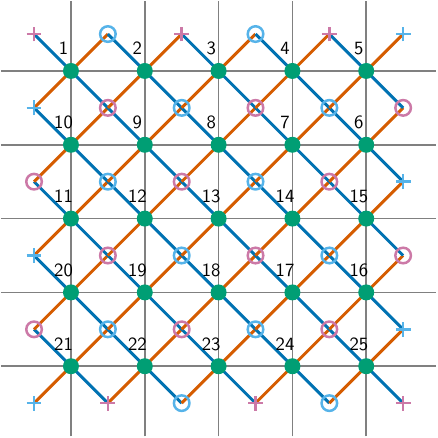}}\par
\subfloat[]{\includegraphics[width=.5\linewidth]{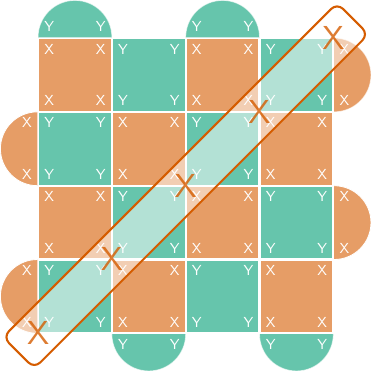}}\hfill
\subfloat[]{\includegraphics[width=.5\linewidth]{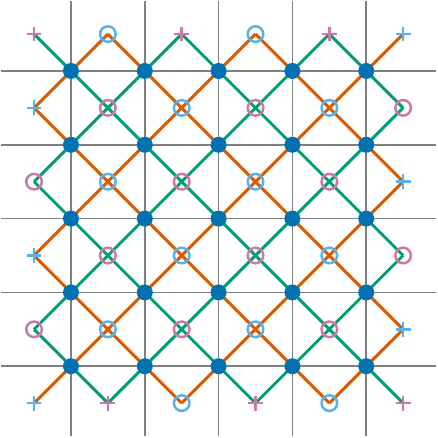}}
\caption{Structure of the XZZX code (a) and XY surface code (c), including two logical operators. The corresponding Ising models (b, d), with two sublattices of Ising spins (pink and purple circles) coupled by intra-lattice two-spin terms and inter-lattice four-spin terms. Couplings $K_x$ (red), $K_y$ (green), and $K_z$ (blue) as given by Eqn.\ \ref{Eq:couplings}. Fixed spins ($+$) are placed along the boundary where the codes have no stabilizers. Vertices are enumerated  $v=1,...,d^2=25$, with the explicit expressions for the two Ising Hamiltonians given by Eqn.\ \ref{Eq:XZZX_H} and \ref{Eq:XY_H}. Quenched disorder, in the form of flipped sign couplings, is not shown.}
\label{fig:fig_lattice}
\end{figure}

The XY code and the XZZX code are modified surface codes tailored to phase-biased noise. We will consider the rotated code geometry, consisting of $d\times d$ qubits, with $d$ an odd integer, and a set of $d^2-1$ parity-check operators as exemplified for $d=5$ in Fig.\ \ref{fig:fig_lattice}. (Generalizing to even $d$ or other geometries is straightforward.)  Since the checks commute they are stabilizers that project an arbitrary density matrix over the $d^2$ qubits into a 2-dimensional subspace, specified by the complete set of $\pm 1$ eigenvalues. Any product of stabilizers is also a stabilizer, but it should be clear from the context if the term refers to the check operators (stabilizer generators) or to operators from the full stabilizer group. The code space, corresponding to the logical qubit, is taken to be the subspace with all $+1$ eigenvalues. A pair of anticommuting operators $X_L$ and $Z_L$, commute with all stabilizers and generate logical bit- and phase-flips respectively. Specifically, we will define $X_L$ as the right diagonal $X^{\otimes d}$ (Fig.\ \ref{fig:fig_lattice}c) for both codes, and $Z_L$ as the left-diagonal $Z^{\otimes d}$ (Fig.\ \ref{fig:fig_lattice}a) for the XZZX code, while for the XY code we define $Z_L$ (not shown) as the operator $Z^{\otimes d^2}$ acting on all qubits. This choice specifies what we refer to as a logical bit-flip ($X_L$) and logical phase-flip ($Z_L$), but all the results are invariant with respect to deformations of the logical operators by any stabilizer. 

We consider standard iid Pauli noise, with error rates $(p_x,p_y,p_z)$ for Pauli-errors $X,Y,Z$ on single qubits, and a total error rate per qubit $p=p_x+p_y+p_z$. An arbitrary ``error chain'' $C\in\{I,X,Y,Z\}^{\otimes d^2}$ thus has a probability 
\begin{equation}
\label{eqn:chain}
\pi_C=(1-p)^{d^2}(\frac{p_x}{1-p})^{n_x}(\frac{p_y}{1-p})^{n_y}(\frac{p_z}{1-p})^{n_z}\,, 
\end{equation}
where $(n_x,n_y,n_z)$ are the number of errors per type. In particular, we will focus on phase-biased noise, specified by $p_x=p_y=p_z/(2\eta)$, where $\eta\geq \frac{1}{2}$ sets the magnitude of the bias.


\subsection{Logical Pauli channel}

To quantify error correction we define the equivalence classes $E(C)\in \{\mathcal{I,X,Z,Y}\}$ of an error chain $C$, according to 
\begin{itemize}
\label{eq:equivalent_class}
\item $\mathcal{I}$: $[C,Z_L]=0$, $[C,X_L]=0$
\item $\mathcal{X}$: $\{C,Z_L\}=0$, $[C,X_L]=0$  (bit-flip)
\item $\mathcal{Z}$: $[C,Z_L]=0$, $\{C,X_L\}=0$  (phase-flip)
\item $\mathcal{Y}$: $\{C,Z_L\}=0$, $\{C,X_L\}=0$ (both)
\end{itemize}
The designation follows from acting on a reference chain $C\in \mathcal{I}$, with a logical bit-flip $X_LC\in \mathcal{X}$, phase-flip $Z_LC\in \mathcal{Z}$, or both $Z_LX_LC\in \mathcal{Y}$. The equivalence class of the chain is invariant under operations by any element of the stabilizer group, as is clear from the fact that the logical operators commute with said group. For a given syndrome, specified by a parent chain $C$, each equivalence class contains $2^{d^2-1}$ error chains.   

A decoder is a map $D:C\rightarrow C'=D(C)$, which given an error chain $C$ outputs a correction chain $C'$, that corresponds to the same syndrome, i.e. it takes the code back to the code space. In practice, since the error chain $C$ is unknown, it is the corresponding syndrome which is decoded. 
Error correction is successful if $E(CC')= \mathcal{I}$. (Since $CC'$ in this case does not have a syndrome it is a stabilizer.) 

A maximum-likelihood decoder (MLD) is an optimal decoder, that outputs a correction chain from the most likely equivalence class. Although exact maximum-likelihood decoding is not feasible beyond the shortest code-distances, there are approximate MLD's based on the Monte-Carlo simulations of error chains using the Metropolis algorithm~\cite{Wootton2012HighCode,Hutter2014EfficientCode,Hammar_2022} or based on tensor network (matrix-product state) representations that allow for efficient evaluation of the class probabilities~\cite{Bravyi2014EfficientCode,qecsim,Chubb2021TensorNetwork}. Recently, deep learning based decoders have also been shown to approach maximum-likelihood accuracy~\cite{lange2023datadriven,varbanov2023neural,bausch2023learning}. In this work we will use a tensor network based decoder~\cite{qecsim}, which allows decoding up to code distances $d\approx 100$ with maintained accuracy and moderate run-times.

Using the fact that the classes
$E(CC')$ are exclusive, we can specify the logical failure rates $P_{\mathcal{X}}$, $P_{\mathcal{Z}}$, and $P_{\mathcal{Y}}$, by summing over all error chains (with probability given by Eqn.\ \ref{eqn:chain}) according to
\begin{equation}
\label{eq:PX}
    P_{\mathcal{X}}=\sum_C\pi_C\delta_{E(CC'),{\mathcal{X}}}
\end{equation}
with $C'=D(C)$ and $\delta$ the Kronecker symbol, and correspondingly for ${\mathcal{Z}}$, and ${\mathcal{Y}}$.  The total failure rate is given by $P_f=P_{\mathcal{X}}+P_{\mathcal{Z}}+P_{\mathcal{Y}}$. The failure rates can also be considered for a particular syndrome in which case the sum in Eqn.\ \ref{eq:PX} is constrained to the corresponding chains, and the probability should be normalized with respect to all four classes. As the error model is agnostic to the state of the code, the logical error channel can be exactly represented by a general Pauli channel 
\begin{align}
\label{Eq:rho}
   \rho' = \epsilon_L(\rho) &= (1-P_f)\rho + P_{\mathcal{X}} X_L\rho X_L+\\
   &P_{\mathcal{Z}} Z_L\rho Z_L+P_{\mathcal{Y}} Z_LX_L\rho X_LZ_L\,,\nonumber
\end{align}
where $\rho'$ is the density matrix of the code after error correction based on the non-erroneous state $\rho$, and $P_{\mathcal{X}}$, etc.\ are given by Eqn.\ \ref{eq:PX}. For numerical simulations the weighted sum of the latter is replaced by a sum over chains sampled according to the distribution, or for a single syndrome by a decoder estimate of the probability of each equivalence class, which may be available using a numerical MLD.

To make a connection to real or simulated experiments on the surface code~\cite{gidney2021stim,lange2023datadriven}, we can also define the non-exclusive bit-flip and phase-flip failure rates $P_{fZ}=P_{\mathcal{X}}+P_{\mathcal{Y}}$ and $P_{fX}=P_{\mathcal{Z}}+P_{\mathcal{Y}}$. These are defined according to whether $CC'$ (the product of error and correction) commutes or not with $Z_L$ and $X_L$ respectively. The first, for example, is what would be measured if the code is repeatedly prepared in an eigenstate of $Z_L$ (i.e. $0_L$ or $1_L$) and after error correction measured in the same basis~\cite{google2023suppressing}. 
To completely characterize the logical failure rate we need the third quantity $P_{fY}=P_{\mathcal{X}}+P_{\mathcal{Z}}$, corresponding to exclusive bit- {\em or} phase-flip errors.  In terms of these, the total logical failure rate can alternatively be expressed as $P_f=\frac{1}{2}(P_{fX}+P_{fZ}+P_{fY})$.
If the logical bit- and phase-flip failure modes are independent, or if an approximate decoder treats them as independent, then  $P_{\mathcal{Y}}=P_{\mathcal{X}}P_{\mathcal{Z}}$, such that $P_f=P_{fX}+P_{fZ}-P_{fX}P_{fZ}$. That the first assumption is not true in general is clear for example from the fact that the optimal threshold for depolarizing noise on the standard surface code is not reached by a decoder that treats the two types of stabilizers independently (see, e.g.\ ~\cite{PhysRevA.108.022401}).

\section{\label{sec:mapping}Mapping to generalized Ising model}
To explore the logical failure rates and thresholds for the XY and XZZX models we will make use of the mapping to generalized Ising models \cite{dennis2002topological,wang2003confinement,PhysRevX.2.021004,kovalev2014spin,chubb2021statistical,dua2022clifford,Vodola2022fundamental,PhysRevLett.131.060603}.  Whereas most studies focus on bulk properties relevant in the thermodynamic limit, here we take care to properly describe the boundaries of the code (see also~\cite{PhysRevLett.131.060603}). The basic idea of the mapping is the following: Consider an error chain $C$. We are interested in calculating the probability of the  equivalence class $E(C)$, $P_{E(C)}\sim\sum_{S}\pi_{S\cdot C}$, where the sum runs over all elements of the stabilizer group. Defining an appropriate Hamiltonian $H_C$ in terms of classical (Ising) spin degrees of freedom $s=\pm 1$, we can represent the class probability as the corresponding partition function $P_{E(C)}\sim\mathcal{Z_C}=\sum_{\{s=\pm 1\}}e^{-H_C(\{s\})}$. Here, and in the rest of text, we have put the inverse temperature $\beta=1$. (This means we we only consider models that satisfy the Nishimori condition which relates quenched disorder to temperature~\cite{nishimori1981internal}) 

The Hamiltonian is defined on the dual to the qubit lattice, with a spin on each plaquette, i.e. a spin for each stabilizer generator. In addition there is a set of fixed spins $s=+1$ along the boundary, in places where a ``missing'' stabilizer would have been triggered by an adjacent error. The Boltzmann weight $e^{-H_C(\{s\})}$ of the state with all $s=+1$ is proportional to the probability of the ``parent'' chain $C$, whereas the weight of a state with certain flipped spins $s=-1$ gives the probability of the chain where the corresponding stabilizers have acted on the parent chain. 
The construction is the following: 
\begin{itemize}
    \item Define coupling constants 
    \begin{align}
      K_x&=-\frac{1}{4}\ln\frac{p_x(1-p)}{p_yp_z}=-\frac{1}{4}\ln\frac{(1-p)}{p_z}\\  
      K_y&=-\frac{1}{4}\ln\frac{p_y(1-p)}{p_xp_z}=-\frac{1}{4}\ln\frac{(1-p)}{p_z}\\ 
      K_z&=-\frac{1}{4}\ln\frac{p_z(1-p)}{p_xp_y}=-\frac{1}{4}\ln\frac{(1-p)4\eta^2}{p_z}\,
      \label{Eq:couplings}
    \end{align}
    where the latter expressions hold for Z-biased noise, with $p_x=p_y=p_z/(2\eta)$. 
    \item For the empty chain $C=I^{\otimes d^2}$ the Hamiltonian is given by connecting spins that correspond to pairs of stabilizers that act with $(Z,Y)$ on a vertex (qubit) with a term $K_xss'$, and that act with $X$ with a term $(K_z,K_y)ss'$. In addition the four spins around each vertex interact through $(K_y,K_z)ss's''s'''$. 
    \item To represent a general parent chain: For a vertex where $C$ has an $X$ error, flip the signs of the couplings $K_z$ and $K_y$. Similarly for a $Y$ or $Z$ flip $K_x,K_z$ or $K_x,K_y$.  The syndrome, as represented by $C$, thus corresponds to $\pm K$ quenched disorder. 
\end{itemize}
With this construction, and the appropriate fixed boundary spins, one can confirm that the relative probabilities of arbitrary error-chains are properly represented. Ferromagnetic order in a Hamiltonian of one of the equivalence classes implies that acting with a logical operator, to change the parent chain, introduces a magnetic domain wall. This has a free energy that scales linearly with the code distance, giving an exponential suppression of the relative class probability, such that the phase transition corresponds to the code-capacity threshold~\cite{dennis2002topological,chubb2021statistical}.  

The Ising Hamiltonians for the empty chain are shown in Fig.\ \ref{fig:fig_lattice}. Enumerating the vertices from top left $v=1,...d^2$ and specifying two sublattices $A$ and $B$ the Hamiltonian for the XZZX code (without disorder) is
\begin{align}
\label{Eq:XZZX_H}
    &H_{XZZX}=\sum_{v\in \text{odd}}H^{o}_v+\sum_{v\in \text{even}}H^{e}_v \\
    &H^{o}_v = K_z s_A s'_A  + K_x s_B s'_B+K_y s_As'_A s_B s'_B \\
    &H^{e}_v=K_z s_Bs'_B+K_xs_As'_A+K_ys_As'_As_Bs'_B 
    \end{align}
and for the XY code
\begin{align}
 \label{Eq:XY_H}
    &H_{\text{XY}}=\sum_{v}H_v \\
    &H_v = K_y s_A s'_A+ K_x s_B s'_B+K_z s_As'_A s_B s'_B \,,
\end{align}
where $s_A$ and $s_A'$ indicate the two spins at vertex $v$ on sublattice $A$, and correspondingly for sublattice $B$.
The XZZX Hamiltonian is equivalent to the zero-field 8-vertex model, whereas the XY Hamiltonian corresponds to the ``staggered'' 8-vertex model, or the Ashkin-Teller model \cite{sutherland1970two,baxter1971eight,PhysRevB.12.429,fan1972critical,PhysRev.64.178,wu1971ising,baxter2016exactly}. 
In the thermodynamic limit (such that the fixed boundary spins can be ignored) there are two separate global $Z_2$ symmetries corresponding to flipping all spins on a sublattice. The two symmetries may thus be broken independently, possibly corresponding to different thresholds for different logical failure modes. For the XZZX code, however, we see that the two sublattices have identical structure, which means that neither or both symmetries will be broken. In contrast, for the XY code, the two sublattices have different coupling constants, and given that the two symmetries are decoupled, they will in general have non-coinciding phase transitions, as is the case for Z-biased noise on the standard `XZ' surface code (see Appendix~\ref{app:xz_surface}). However, for the special case $p_x=p_y$, i.e.\ $K_x=K_y$, considered here, we see that the two sublattices are identical, and will give a single threshold.

\section{\label{sec:exact}Exact solutions at special disordered point}

In the following we show that the two models can be solved exactly, for arbitrary code distance, for Z-biased noise $p_x=p_y=p_z/(2\eta)$, at the special point $p_s=\frac{1+\eta^{-1}}{2+\eta^{-1}}$, corresponding to $p_z=(1-p)$. Here, according to Eqn.\ \ref{Eq:couplings}, $K_x=K_y=0$, such that only the $K_z$ coupling is finite. We will see that both models and for all four equivalence classes can be reduced to one-dimensional Ising models, from which we can calculate the relative probabilities of the classes. In addition, we can show that this holds equivalently for any syndrome, which means that we can calculate the exact expressions for the logical failure rates $P_\mathcal{X}$, $P_\mathcal{Y}$, and $P_\mathcal{Z}$, under maximum-likelihood decoding.   

\subsection{XZZX code at $p=p_s$}

\begin{figure}
\includegraphics[width=0.9\linewidth]{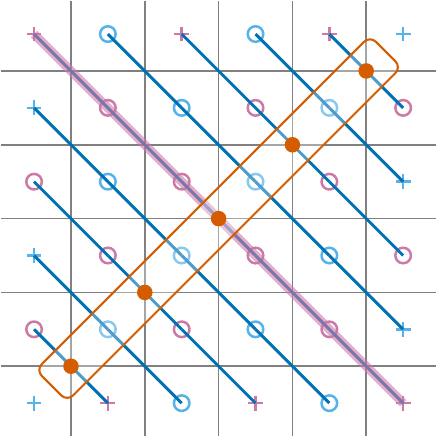}
\caption{XZZX code for $K_x=K_y=0$, and mapping to a set of decoupled 1D Ising models, exemplified for $d=5$. The logical operator $X_L$ (red dots) flips the signs of the couplings at the corresponding vertices. Only the central left diagonal Ising model has fixed spins at both ends, which implies that only for this is the free energy affected by the sign change.}
\label{fig:Z-biased}
\end{figure}

The XZZX model directly reduces to a set of decoupled one-dimensional, $\pm K_z$ random-bond Ising models, as shown in Fig.\ \ref{fig:Z-biased}. The 1D Ising model can be solved exactly for arbitrary couplings, but to avoid having to deal with the quenched disorder we make the following observation: Any syndrome in the XZZX model can be represented by an error chain containing only Pauli $Z$ and $I$. There are $2^{d^2}$ unique chains containing only $Z$ and $I$. The only pure Z operator which does not change the syndrome, is the diagonal $Z_L$, with weight $d_Z=d$, which means that the pure $Z/I$ chains come in pairs with the same syndrome. These then specify $2^{d^2}/2=2^{(d^2-1)}$ unique syndromes, which is the complete set of syndromes. Clearly, the pair of such chains are in class $\mathcal{I}$ and $\mathcal{Z}$, and the following analysis holds equivalently for any syndrome, corresponding to such a pair of chains. As $Z$ flips the signs of the $K_x$ and $K_y$ terms, which are both zero at the special point, the Hamiltonians $H_{C_{\mathcal{I}}}$ and $H_{C_{\mathcal{Z}}}$ will be identical, both corresponding to decoupled one-dimensional Ising models, without any flipped sign disorder of the coupling $K_z$. Consequently the free energies are the same, and $P_\mathcal{I}=P_\mathcal{Z}$. As these are 1D Ising models, they are disordered, except at $\eta\rightarrow\infty$, where $K_z\rightarrow -\infty$, such that the spins are ferromagnetically ordered.  The point  $p=\frac{1+\eta^{-1}}{2+\eta^{-1}}\geq 0.5$ is thus (as expected) above the threshold, except in the limit of pure Z-biased noise, where the point $p=0.5$ is at the threshold. 

In the thermodynamic limit, given that there is a unique threshold by symmetry of the two sublattices, all four classes must be equally probable. However, with the explicit representation of the boundary of the code we can also solve for finite size effects. Starting with the chain $C_{\mathcal{I}}$, we act with $X_L$, using the pure-$X$ representation along the diagonal, to get a parent chain $C_{\mathcal{X}}$. Vertices along the diagonal will change from a $Z$ or $I$ to a $Y$ or $X$ respectively, which flips the sign of $K_z$ and (irrelevant at the special point) $K_x$ or $K_y$ at the vertex. The Hamiltonian $H_{C_{\mathcal{X}}}$ is again a set of decoupled 1D Ising models, but with one flipped sign coupling on every other diagonal (Fig.\ \ref{fig:Z-biased}). Using the standard transfer matrix method we can calculate the partition function of the Ising models, with the total being a product of decoupled models. Only the center diagonal Ising model, with fixed spins at both ends, is sensitive to the sign change, and as shown in Appendix \ref{app:one-d-ising} the relative probability of the classes is given by
\begin{equation}
\label{eq:class_prob_Z}
   \frac{P_\mathcal{X}}{P_\mathcal{I}}=\tanh(d_Z\,\text{artanh}(1/2\eta))\approx\tanh{\frac{d_Z}{2\eta}}
\end{equation}

with $d_Z=d$, and where the last expression holds for $\eta\gg 1$. The final class, $\mathcal{Y}$, is given by acting with $Z_L$ on the parent chain of class $\mathcal{X}$. As this only flips the signs of vanishing couplings, the corresponding Hamiltonians are the same, giving  $P_\mathcal{Y}=P_\mathcal{X}$. From Eqn.\ \ref{eq:class_prob_Z} we see that the finite size effects can be quite severe, with the probability of logical bit-flips being suppressed. To approach the thermodynamic limit, corresponding to $P_\mathcal{X}=P_\mathcal{I}$, requires $d> \eta$.  

Given the class probabilities, and that they are the same for all syndromes, we can calculate the full logical failure rates. Decoding is trivial: since $P_\mathcal{I}=P_\mathcal{Z}\geq P_\mathcal{X}=P_\mathcal{Y}$, optimal decoding is accomplished by providing a pure-$Z$ chains that correspond to the syndrome. (The latter can be found by individually generating each syndrome defect using a string of $Z$'s connecting to a boundary where it is not detected by a boundary stabilizer.)  
The failure rates are given by 
\begin{align}
\label{Eq:failure_XZZX_special_1}
    P_{fX}&=P_{fY}=\frac{1}{2}\\
    \label{Eq:failure_XZZX_special_2}
    P_{fZ}&=\frac{1}{2}-\frac{1}{2}e^{-2d_Z\,\text{artanh}(1/2\eta)}\approx \frac{1}{2}-\frac{1}{2}e^{-d_Z/\eta}\\
    \label{Eq:failure_XZZX_special_3}
    P_{f}&=\frac{3}{4}-\frac{1}{4}e^{-2d_Z\,\text{artanh}(1/2\eta)}\approx \frac{3}{4}-\frac{1}{4}e^{-d_Z/\eta}  
\end{align}
where the final expressions hold for $\eta\gg 1$. We thus find that the rate of logical bit-flip errors is strongly suppressed for code distances $d<\eta$, which also affects the total logical failure rate. Although the special point is above the threshold for the XZZX code (except for $\eta=\infty$), it influences, as we will confirm numerically, the failure rates far below the threshold, in the form of large finite size corrections.
Note that for depolarizing noise $\eta=1/2$, the special point corresponds to the trivially disordered point $p_x=p_y=p_z=1-p$, with $p=0.75$, where finite size corrections, according to Eqn. \ref{Eq:failure_XZZX_special_2} and \ref{Eq:failure_XZZX_special_3}, also vanish.

\subsection{XY code at $p=p_s$}

The analysis of the XY model at the special point $p_x=p_y$, and $p_z=1-p$ follows the same outline as for the XZZX model. We again define $Z_L$ as the unique pure Z operator, which now acts on all qubits, with weight $d_Z=d^2$. Thus, any syndrome can be specified by a chain which contains only $Z$ and $I$, with the parent chain $C_\mathcal{I}$ having an even number of $Z$'s on the right diagonal (thus commuting with $X_L$) and $C_\mathcal{Z}=Z_LC_\mathcal{I}$ having an odd number. The Hamiltonians $H_{C_\mathcal{I}}=H_{C_\mathcal{Z}}$ are uniform, containing only the four spin interactions with coupling constant $K_z$.

\begin{figure}
\includegraphics[width=0.9\linewidth]{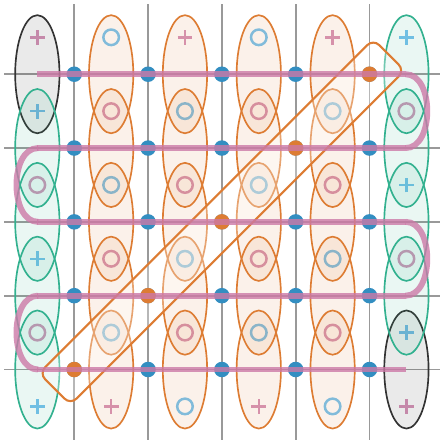}
\caption{XY code for $K_x=K_y=0$ and mapping to 1D Ising model, following the marked path, exemplified for $d=5$. Pairings show new composite Ising spin variables. Note that the due to the fixed boundary spins, subsequent composite spins (green) on the boundary are identical, and the initial and final spins (black) are fixed. Also shown is the logical operator $X_L$ (red dots) which flips the sign of the couplings on the corresponding vertices.}
\label{fig:Y-biased}
\end{figure}

These Hamiltonians can be solved by defining new Ising variables $\sigma=s_As_B$ on each horizontal bond, see Fig.\ \ref{fig:Y-biased}, which gives a mapping to a single one-dimensional Ising model with terms $K_z\sigma\sigma'$ that follows the path of enumerated vertices (see Fig.\ \ref{fig:fig_lattice}) over all $d^2$ vertices. This simple result follows from the fact that the composite spins at the turning points of the string, containing one fixed and one free spin, are in fact the same.  
In addition, the configuration of fixed spins at the boundary implies that the Ising model has fixed spins $\sigma=+1$ at both ends. The operator $X_L$ flips all couplings along the right diagonal and the relative class probabilites follow from the exact solution to the partition functions. The results are identical to the those of the XZZX model, Eqn.\ \ref{Eq:failure_XZZX_special_1}-\ref{Eq:failure_XZZX_special_3}, with the modification that the pure noise code-distance is now given by $d_Z=d^2$. 
As for the XZZX code, the special point is a disordered point, i.e.\ above the threshold, except at $\eta\rightarrow\infty$. We see that the logical failure rate has significant finite size corrections, although not as severe as for the XZZX code. For the XY code, code distance $d>\sqrt{\eta}$ is sufficient to approach the thermodynamic limit.

\begin{figure}\centering
\subfloat[]{\includegraphics[width=\linewidth, trim={0.7cm 0.2cm 0.6cm 1cm},clip]{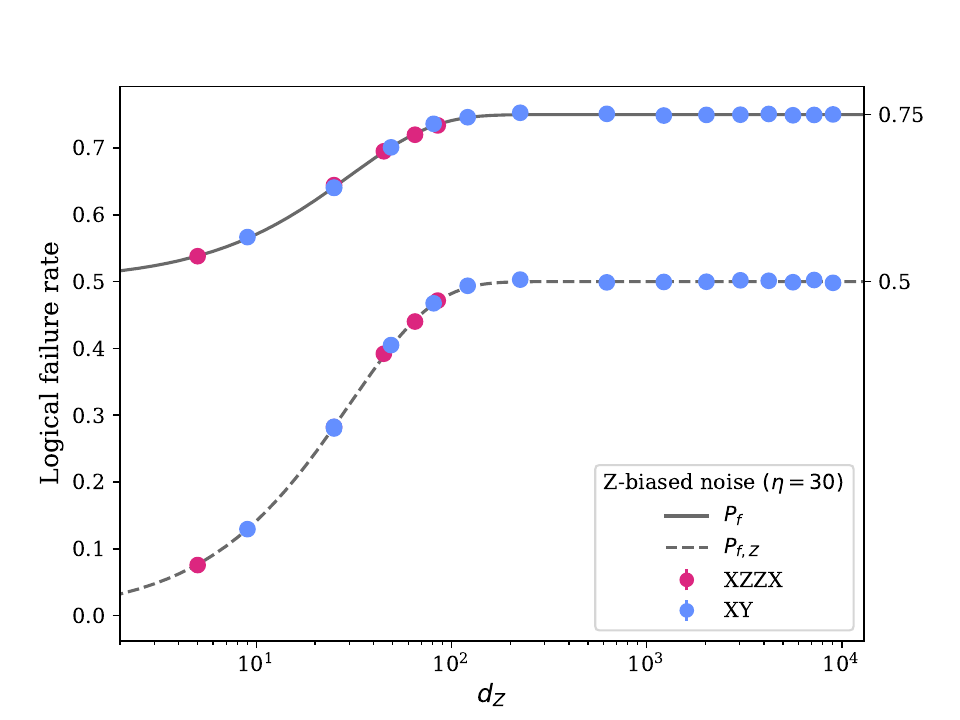}}\par\vspace{-0.1cm}
\subfloat[]{\includegraphics[width=\linewidth, trim={0.7cm 0.2cm 0.6cm 1cm},clip]{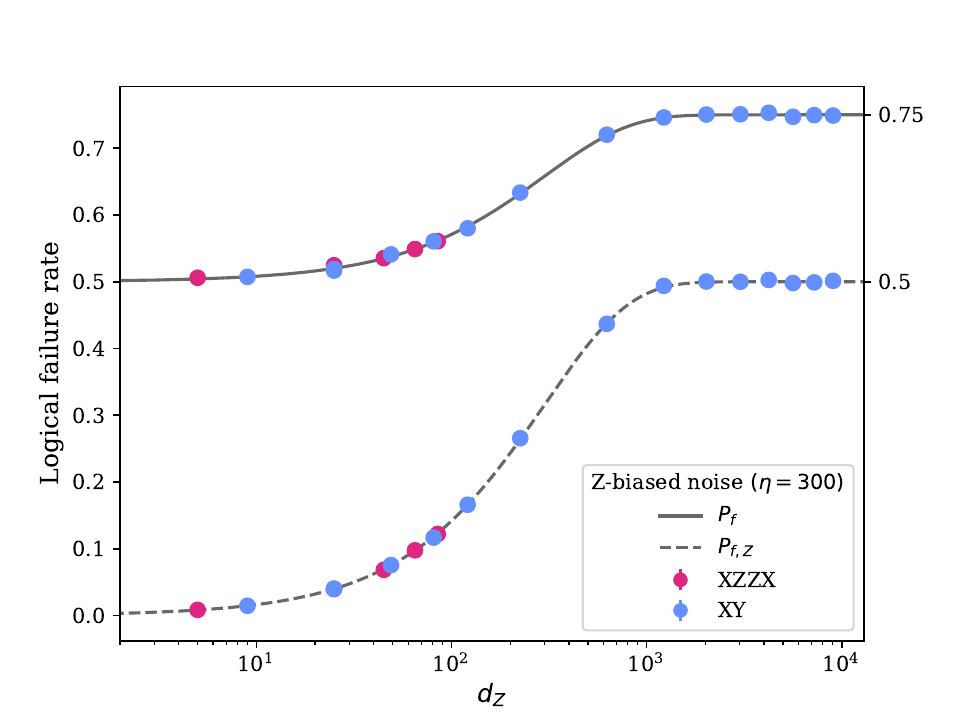}}
\caption{Total ($P_f$) and bit-flip ($P_{fZ}$) logical failure rates at the error rate $p_s=\frac{1+\eta^{-1}}{2+\eta^{-1}}$ versus effective code distance $d_z=d$ for the XZZX code and $d_z=d^2$ for the XY code. Lines correspond to exact solutions, Eqn.\ \ref{Eq:failure_XZZX_special_2}-\ref{Eq:failure_XZZX_special_3}, and data-points are numerical results. The latter are averaged over 60,000 syndromes, but for each syndrome the failure rates are the same up to decoder accuracy, in accordance with the exact solution.}
\label{fig:exact_results}
\end{figure}

Figure \ref{fig:exact_results} shows a comparison between the exact results, Eqn.\ \ref{Eq:failure_XZZX_special_2}-\ref{Eq:failure_XZZX_special_3}, and numerical calculations of the failure rates up to code distances $d\approx 100$. The agreement confirms the accuracy of the numerical decoder, while showing that operable code distances correspond to a large finite size suppression for the XZZX code.   

\subsection{Order of limits for pure noise}
For pure Z noise, the failure rate of both codes can be calculated exactly for any error rate and any code distance~\cite{Ataides2021XZZX,Srivastava2022xyzhexagonal}. As there is only the single pure Z logical operator, $Z_L$, and syndromes are uniquely specified by the corresponding pair of chains, the code is equivalent to a distance $d_Z$ classical repetition code which has a logical failure rate $P_f=0.5$ at $p=0.5$. (There are no logical bit-flip errors.) From the exact solution Eqn.\ \ref{Eq:failure_XZZX_special_1}-\ref{Eq:failure_XZZX_special_3}, however, we note that the order of the limits, $\eta\rightarrow\infty$ and $d\rightarrow\infty$, do not commute. Taking the limit $d\rightarrow\infty$ first gives a total logical failure rate of $P_f=0.75$ at the point $p=0.5$, corresponding to both logical bit and phase disorder. The opposite limit gives the total logical failure rate of $P_f=0.5$, corresponding to a pure logical phase-flip channel. This is in a sense an extreme case of finite size correction to the failure rate at threshold, although here it does not imply any uncertainty in the value of the threshold.

\section{\label{sec:results}Threshold estimates}
Using the exact results as a reference we extend the analysis to lower error rates using a numerical decoder, with the aim to clarify the influence of finite size effects on the estimated thresholds for varying bias. The tensor network based decoder of \cite{qecsim} was used, which is based on the original formulation in \cite{Bravyi2014EfficientCode}. We have modified the decoder to separately keep track of failures with respect to three logical Pauli operators. A bond-dimension of $\chi=16$ for $\eta< 30$ and $\chi=8$ for $\eta\geq 30$ is used, which is sufficient for good convergence for the XZZX code for considered code distances, as shown in~\cite{Ataides2021XZZX}. For the XY code, we find that convergence of the failure rate with the bond dimension is significantly worse, for error rates around the threshold. Given that the finite-size effects for the XY model are smaller, and that the thresholds are already convincingly found to be very close to the hashing bound, using approximate self-duality~\cite{dua2022clifford}, we only present further results for the XZZX model.

\begin{figure}\centering
\subfloat[]{\includegraphics[width=\linewidth, trim={0.5cm 0cm 1cm 1cm},clip]{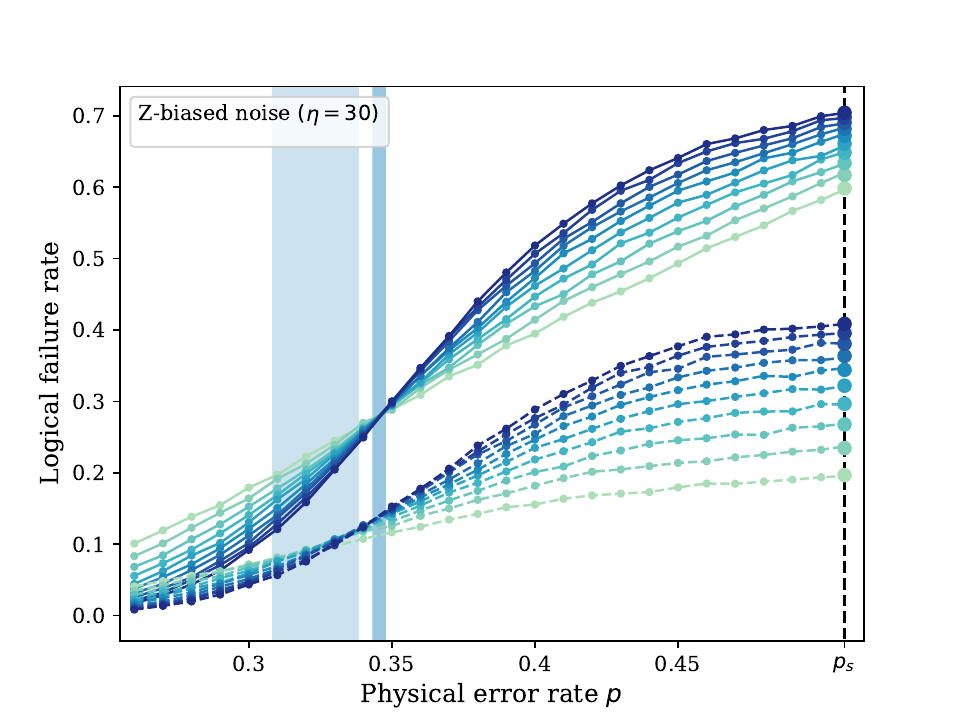}}\par\vspace{-0.3cm}
\subfloat[]{\includegraphics[width=\linewidth, trim={0.5cm 0cm 1cm 1cm},clip]{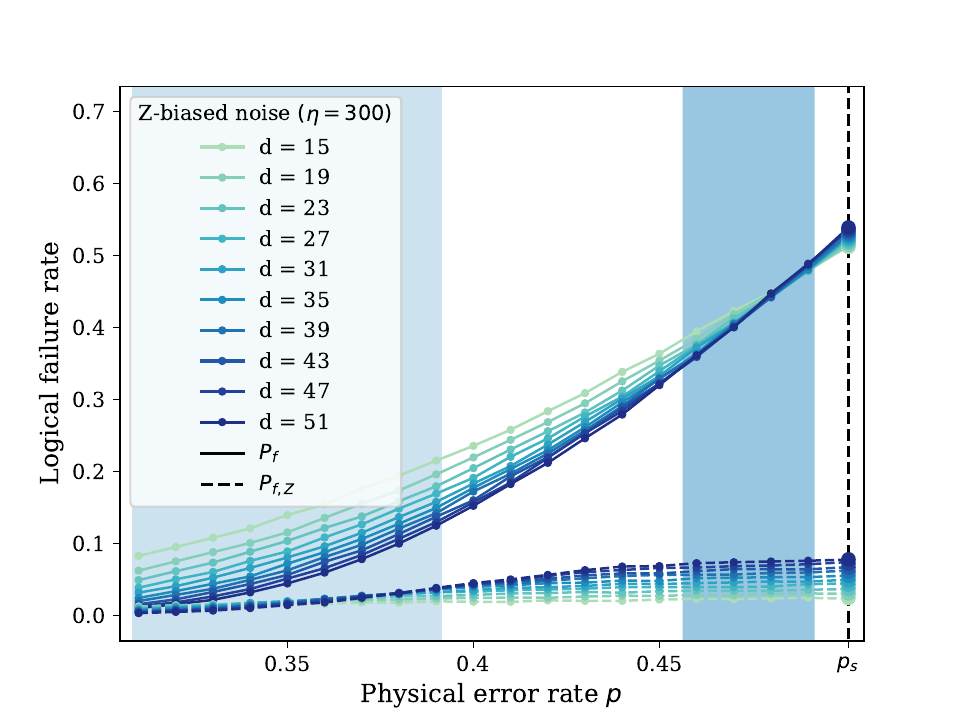}}
\caption{Total ($P_f$) and bit-flip ($P_{fZ}$) logical failure rates for moderate ($\eta=30$) (a) and large ($\eta=300$) (b) bias, versus error rate $p$, for the XZZX code, illustrating the discrepancy between thresholds extracted from the two measures, due to finite size effects. Shaded regions indicate the approximate range of threshold values for the range of code-distances plotted, as discussed in Sec.\ \ref{sec:fits}. Sampled over 60,000 randomly generated syndromes decoded using the tensor-network decoder~\cite{qecsim}. Points at $p_s=\frac{1+\eta^{-1}}{2+\eta^{-1}}$ correspond to exact values using Eqn.\ \ref{Eq:failure_XZZX_special_2}-\ref{Eq:failure_XZZX_special_3}.}
\label{fig:failure_rates}
\end{figure}

To set the stage, Fig.\ \ref{fig:failure_rates} shows the total logical failure rate $P_f$ and the bit-flip failure rate $P_{fZ}$ for two different bias and varying code-distances over a range of error rates $p$ that clearly contain the thresholds. We see that the finite size correction at the point $p=p_s$ also extends to significantly lower error rates, which, as discussed in the next section, complicates the accurate extraction.   


\subsection{Finite size scaling fits}
\label{sec:fits}
To estimate thresholds from numerical data it is customary to fit failure rates to a finite size scaling form, motivated by the correspondence to the phase transition of the generalized Ising model~\cite{wang2003confinement,harrington2004analysis}. Close to the transition $p=p_c$ the correlation length $\xi$ (corresponding to the typical size of a magnetic domain) diverges as $\xi\sim 1/(p-p_c)^\nu$, where $\nu$ is a critical exponent. In the regime where the linear dimension of the code is effectively small, $d\ll \xi$, by scale invariance, the properties of the code are expected to be controlled by the ratio $d/\xi$, such that $P_f\sim f(d/\xi)$ where $f$ is a scaling function (see, e.g.\ \cite{goldenfeld2018lectures}). Re-expressing this in terms of the quantity $x=(p-p_c)d^{1/\nu}$ and expanding for small $x$ gives the standard fitting form 
\begin{equation}
\label{eq:fit2}
   P_f=A+Bx+Cx^2\, . 
\end{equation}
Importantly, this form implies that the failure rates should (within the accuracy of the data) cross at the threshold. This, we find, is not a viable ansatz for the data in Fig. \ref{fig:failure_rates}. Indeed, the scaling form ignores the explicit influence of the boundaries, which give finite size corrections $g(d,p)$, with $g(d,p)\rightarrow 0$ as $d\rightarrow \infty$. Even though we know the explicit expression for the finite size correction for the total failure rate $g(d,p=p_s)\approx -\frac{1}{2}e^{-d/\eta}$ we have found that it is difficult to extract the correction close to the threshold, given the statistical fluctuations that influence the quality of the data. Instead we follow the standard recipe, and use the fitting formula (without finite size correction) for different bins of progressively larger code distances to extract $d$-dependent estimates of the threshold~\cite{Ataides2021XZZX}. Different from the standard approach, we  do this independently for failures with respect to the three different logical operators, $P_{fX}$, $P_{fY}$, and $P_{fZ}$, and the total failure rate $P_f$. Given that we know that there is a unique threshold, convergence with respect to code distance for all four (three independent) measures gives a high degree of confidence in the threshold estimate. Results from this analysis for the XZZX code is shown in Figure~\ref{fig:thresholds} and Appendix \ref{app:thresholds}. Examples of the fits are also shown in Appendix \ref{app:thresholds}. We find that for $d\ll \eta$ the total logical failure rate overestimates the threshold, whereas the bit-flip failure rate underestimates the threshold. For moderate $\eta$, operable code distances are large enough to find convergence, giving a very confident estimate of the threshold. Figure \ref{fig:threshold_vs_eta} summarizes these findings, showing well converged estimates of the thresholds for $\eta\lesssim d_{max}$, where $d_{max}=101$  are the largest code distances used in this study.

\begin{figure}\centering
\subfloat[]{\includegraphics[width=\linewidth, trim={0.4cm 0.2cm 1.6cm 1cm},clip]{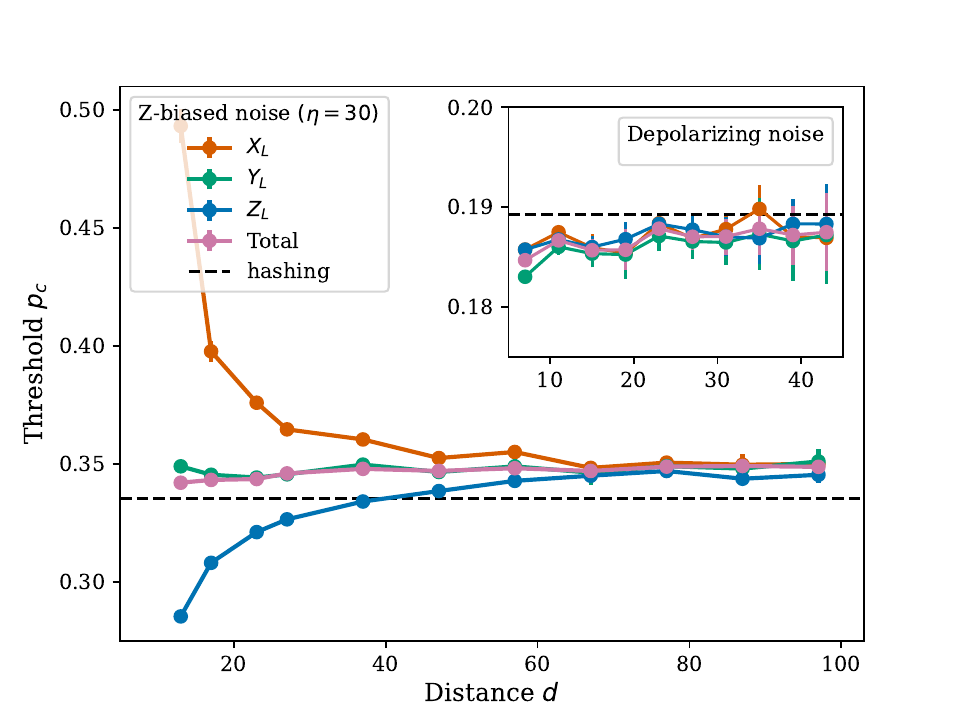}}\par\vspace{-0.3cm}
\subfloat[]{\includegraphics[width=\linewidth, trim={0.4cm 0.2cm 1.6cm 1cm},clip]{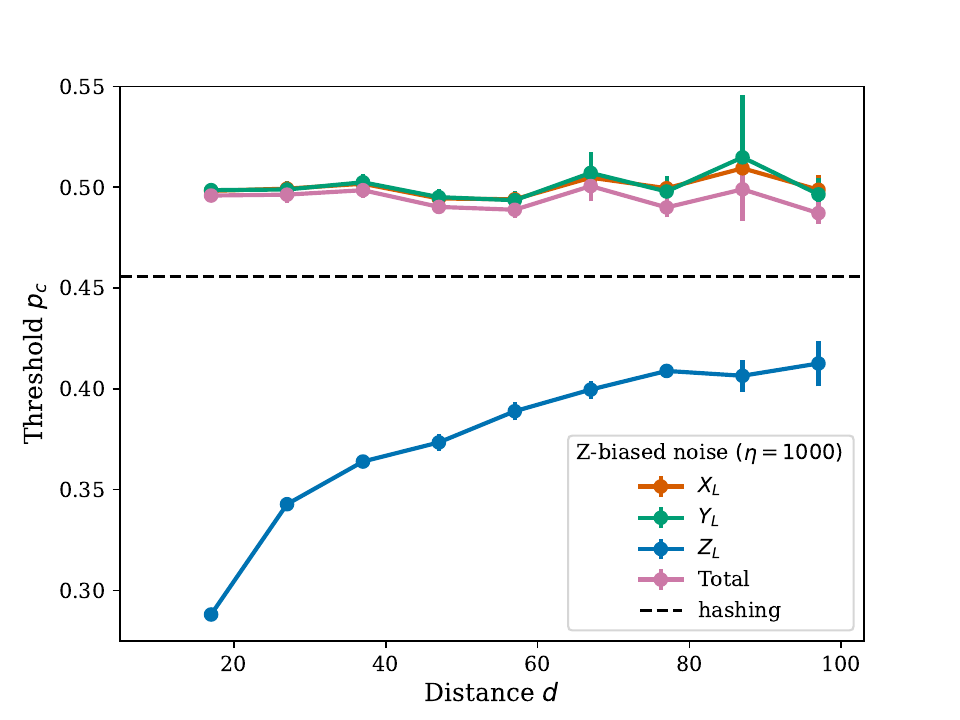}}
\caption{Estimated threshold values $p_c$, versus code distance $d$, for the four logical failure rates, $P_{fZ}$ (bit-flip), $P_{fX}$ (phase-flip), $P_{fY}$, and $P_{f}$ (total), for the XZZX code. In the thermodynamic limit these should converge to a single threshold value, as seen for $\eta=30$ (a). For $\eta=1000$, numerically accessible code distances are too small for convergence (b).  Inset shows depolarizing noise $\eta=1/2$, for which no systematic code-distance dependence is evident. Dashed lines indicate the hashing bound. Error bars correspond to uncertainties in the fit, as discussed in Appendix \ref{app:thresholds}.}
\label{fig:thresholds}
\end{figure}



\begin{figure}
\includegraphics[width=\linewidth, trim={0.2cm 0cm 1.6cm 1.4cm},clip]{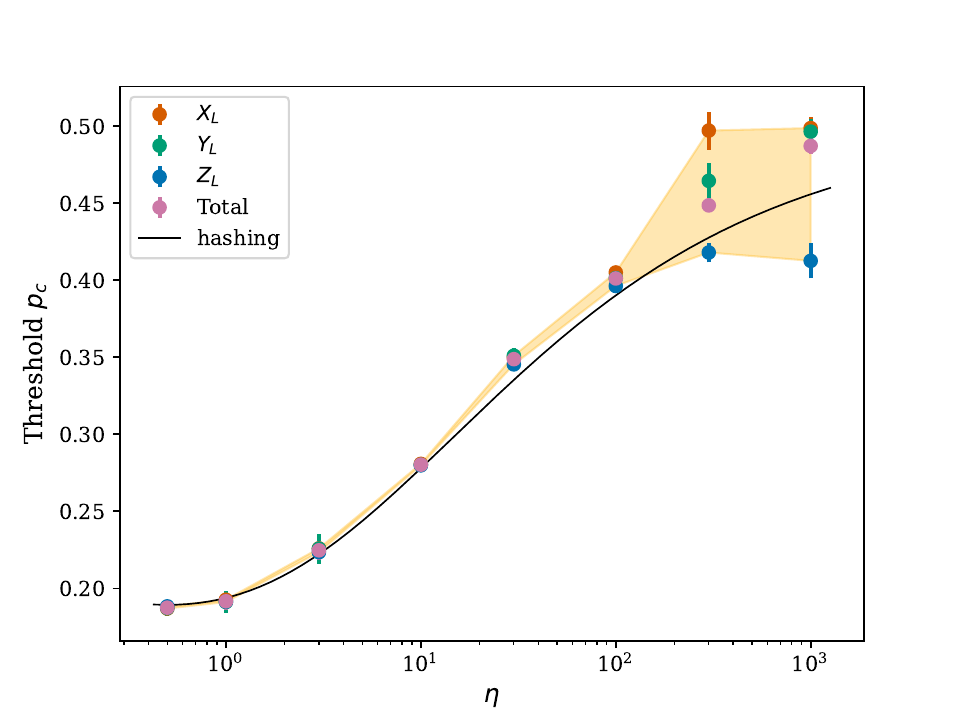}
\caption{Estimated threshold $p_c$, versus noise bias $\eta$, for the XZZX code. Data corresponds to the threshold fits for the largest code-distance from Fig. \ref{fig:thresholds}, \ref{fig:eta100}, and \ref{fig:eta300}. For intermediate $\eta$ the convergence of the measures confirms that the  threshold is above the hashing bound, whereas for larger $\eta$ the lack of convergence (shaded region) makes the results inconclusive.}
\label{fig:threshold_vs_eta}
\end{figure}

\section{Low-weight logical bit-flip operators}
\label{sec:low_weight}
\begin{figure}
\includegraphics[width=\linewidth, trim={0.15cm 0cm 1.6cm 1.2cm},clip]{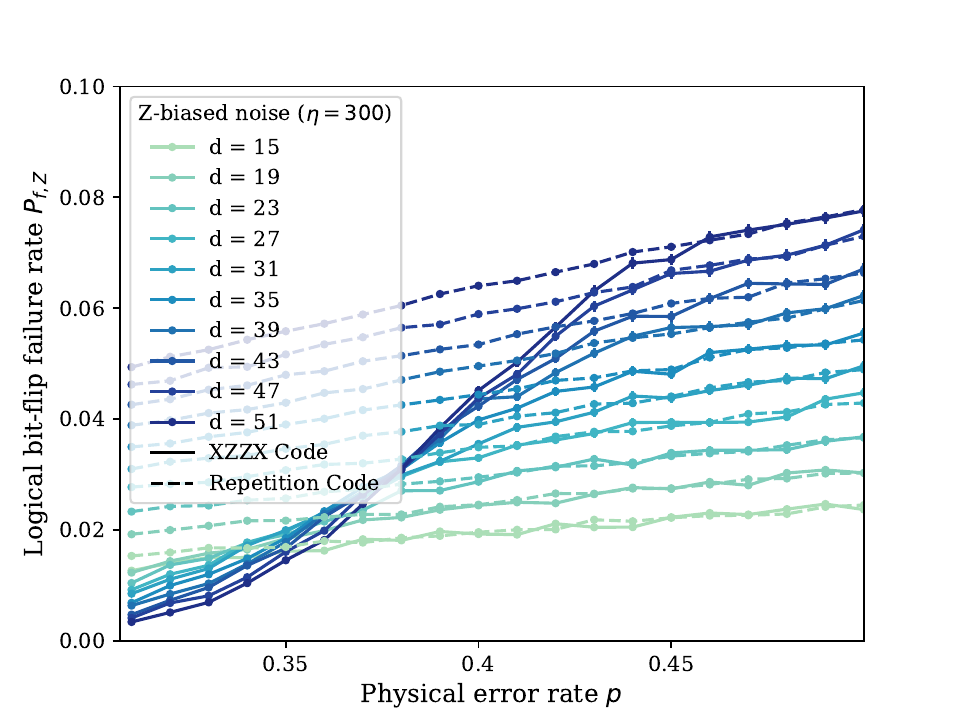}
\caption{Logical bit-flip failure rate versus error rate for code-distance $d$ much less than the phase-bias $\eta$. Comparing the XZZX code with $d\times d$ data qubits to the phase error correcting repetition code with $d$ data qubits. Above a $d$-dependent cross-over region the two codes have very similar logical failure rates, showing that the former has no fault tolerance to bit-flip errors.}
\label{fig:cf_repetition_code}
\end{figure}

To provide a simple explanation for why the logical bit-flip failure rate is suppressed close to the threshold for $d\ll \eta$, we can use the following explicit constructing of logical $X_L$ operators containing a single minority, $X$ or $Y$, error. For the XZZX code, consider an arbitrary error chain $C$ containing any number of $Z$'s together with a single $X$ or $Y$ placed on any of the $d_z=d$ qubits on the left diagonal, i.e.\ coinciding with the pure-$Z$ logical, $Z_L$. As shown in Sec. \ref{sec:exact}A, any syndrome of the XZZX code can be represented by two pure $Z$-chains, $C'$ and $Z_LC'$. Thus, picking such a chain $C'$ with the same syndrome as $C$ and satisfying $[CC',X_L]=0$, the product $CC'$ will be a logical operator $X_L$, since $\{CC',Z_L\}=0$. This set of logical bit-flip operators containing a single $X$ or $Y$ gives the dominant contribution to the bit-flip failure rate $P_{fZ}$ at high overall error rates (where $Z$-errors are cheap) for code sizes such that $d_zp_x\approx d_z/\eta \ll 1$, corresponding to the lowest order non-vanishing term in an expansion of Eqn.\ \ref{Eq:failure_XZZX_special_2}. Interestingly, as the logical failure rate is dominated by single bit-flip errors in this regime it implies that the effective code distance is $d_x=1$, there is no protection against bit-flip errors. This is demonstrated by Fig.\ \ref{fig:cf_repetition_code} that compares the logical bit-flip failure rate of the XZZX code to the phase-flip detecting repetition code\cite{PhysRevX.9.041053,PRXQuantum.3.010329} with the same code-distance $d$ and the same biased noise model. There is a code-distance dependent cross-over, between a high error rate (near-threshold) regime where the XZZX model is effectively equivalent to the repetition code, to a low error rate regime where the full code-distance is manifested. Thus, in this near-threshold regime and considering code-distances $d\ll \eta$, the XZZX code is inefficient compared to the repetition code, as both have phase-flip error correction corresponding to code distance $d_z=d$, while the full matrix of $d^2$ data qubits of the former does not provide any additional bit-flip tolerance.   

\begin{figure}
\includegraphics[width=\linewidth, trim={0cm 0cm 0cm 0cm}, clip]{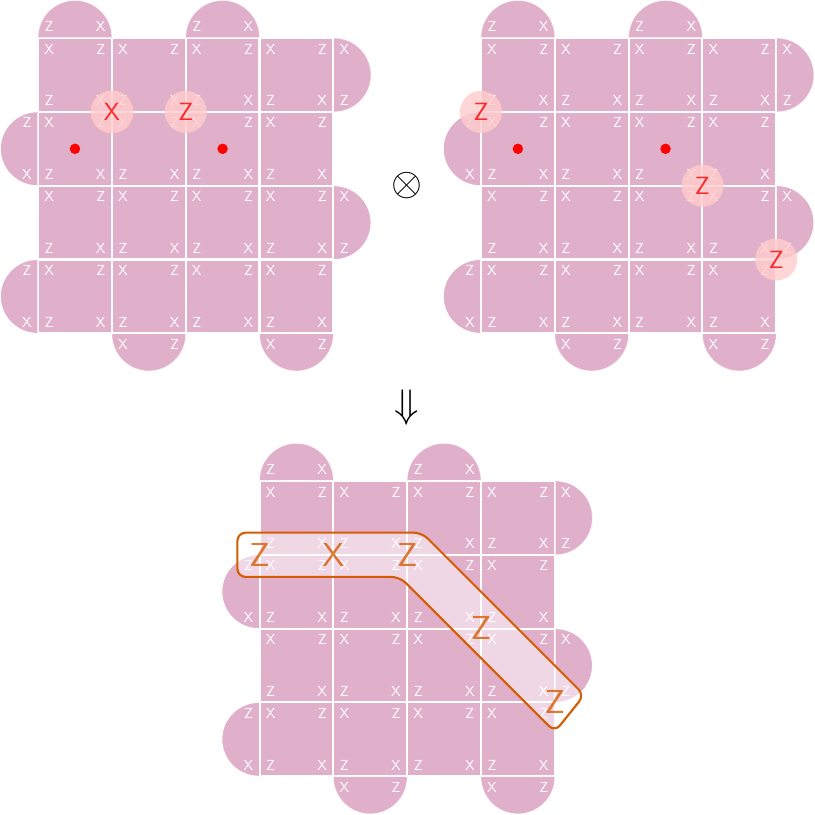}
\caption{Construction of logical bit-flip operator, containing a single minority  error for the XZZX code, as the product of any error chain with a single $X$ or $Y$ on the left diagonal and a pure-$Z$ chain with the same syndrome. }
\label{fig:single-x-logical}
\end{figure}

The construction of low minority-weight logical bit-flip operators is equivalent for the XY code, but where a chain with a single $X$ or $Y$ on any of $d_z=d^2$ qubit is  multiplied by a pure-$Z$ chain with the same syndrome (see \figref{fig:single-x-logical}). 
Note that this construction of logical bit-flip operators on the XY code is distinct from the operators discussed in terms of ``fragile boundaries'' \cite{PhysRevX.13.031007}. The latter refer to logical {\em phase-flip} operators that contain a single $X$ or $Y$ and order $d$ $Z$'s which in low  error rate regime may dominate over the weight $d^2$ pure $Z$ operator even at large bias.    

The construction of logical bit-flip operators, with a single minority error, giving a lowest order bit-flip failure rate $P_{fZ}\sim d_z/\eta$ at high error rates, applies also to more general Clifford-deformed\cite{dua2022clifford} codes with a weight $d_z$ pure-$Z$ logical operator. However, for codes that also have pure-$Z$ stabilizers the construction would have to be modified, and might not be valid, as not all syndromes can be represented by pure $Z$-chains. In particular, for the standard CSS surface code the construction does not hold. The effective $d_x$ code-distance cannot be reduced by transforming low-probability to high-probability errors through stabilizers.     

\section{\label{sec:discussion}Discussion}


From our findings we make the following observation concerning the code capacity thresholds for the XZZX code under highly biased noise. Ataides {\em et al.} \cite{Ataides2021XZZX} were careful to systematically explore the convergence with respect to code distance, however, only considering the total failure rate $P_f$. It was found that that the threshold at $\eta=300$ was not converged with respect to code distance, while the threshold for $\eta=1000$ was. As we see in our study, convergence of the threshold using the total failure rate is not a guarantee for correctness. In fact, from the exact results at $p=p_s$, the available code distances are much too small to resolve the thermodynamic limit for $\eta=1000$, as the contribution from logical bit-flip failures is almost completely suppressed. The conclusion from our study is that the threshold values cannot be deduced with confidence from numerical decoder data at $d\ll \eta$. Nevertheless, our results confirm the main conclusion in \cite{Ataides2021XZZX} concerning the code capacity thresholds, namely that for some range of bias they are indeed above the hashing bound. In fact, as found in \cite{dua2022clifford}, the XZZX and the XY code can be thought of as special periodic codes in a broader family of Clifford-deformed random codes with thresholds that follow closely the hashing bound. However, also in \cite{dua2022clifford} we would argue that the deduced threshold values are very uncertain for large bias, due to large finite size effects. The latter is clear from the fact that, for the code sizes considered, $P_f$ reaches a maximum near $0.5$ instead of $0.75$ around an error rate $p=0.5$, which is certainly above the threshold for any finite bias. For such random Clifford deformed codes with high thresholds, the point $p=p_s$ corresponds to a percolating cluster of two- and four-term Ising interactions. The finite size properties of the failure rates of such clusters will be interesting to explore in future work, and as discussed in Sec.\ref{sec:exact}D, whether an effective pure-noise code distance $d_z$ will also characterize those. 

Finally, we comment on surface codes realizations that do not have a single threshold for phase-biased noise, exemplified by the standard rotated `XZ' surface code. At $p=p_s$ this maps to two decoupled two-dimensional random-bond Ising models, with coupling constants $K_z\neq 0$ and $K_x=0$. Since acting with $X_L/Z_L$ flips the sign of $K_z/K_x$ the former couples to logical bit-flip errors and the latter to logical phase-flip errors. The latter is trivially disordered, corresponding to a logical phase-flip failure rate $P_{fX}=0.5$ at $p=p_s$ independently of code distance. The former (see Appendix \ref{app:xz_surface}) corresponds to the random bond Ising model which represents a pure-noise channel on the surface code, with a low effective error rate, such that even for moderate bias ($\eta\gtrsim 4$) the threshold to logical bit-flip errors is above $p=p_s\gtrsim 0.5$. 
The finite size correction for large error rates $p\approx 0.5$ thus has the opposite sign to that of the XZZX and XY codes, with total failure rate $P_f\rightarrow 0.5^+$, but now with the standard sub-threshold exponential scaling with code distance.

\section{\label{sec:conclusion}Conclusion}

In conclusion, we find exact solutions for arbitrary code distances to the logical failure rates of the XY and the XZZX models under phase-biased noise $p_x=p_y=p_z/2\eta$, at a special point $p=p_s=\frac{1+\eta^{-1}}{2+\eta^{-1}}$, under standard ideal conditions. The solutions follow from mapping the codes to generalized random bond Ising models, that at the special point reduce to one-dimensional Ising models, from which the relative probability of the four logical equivalence classes can be calculated exactly and equivalently for any syndrome. In the thermodynamic limit the special point is in the disordered phase of the Ising model, i.e.\ above the code-capacity threshold, except for pure phase biased noise ($\eta\rightarrow\infty$) where it coincides with the threshold. 

From the exact solutions we find that the logical bit-flip failure rate, for large bias, $\eta\gg 1$, is suppressed by a factor $1-e^{-d_z/\eta}$, where $d_z$ is the effective code distance for pure phase-noise, and with a corresponding reduction of the total logical failure rate. This implies a significant finite size effect that influences the reliability of threshold estimates based on finite-size-scaling fits to numerically calculated failure rates for moderate code distances. We demonstrate this by studying the convergence with code distance of threshold fits for failures with respect to the three logical Paulis separately, compared to only fitting to the total failure rate. For the XZZX model this confirms that the thresholds for moderate bias, $\eta=30$ and $\eta=100$, are above the hashing bound, whereas for larger bias, the finite size corrections are too large to determine precise thresholds. Based on this analysis, we also identify a regime close to the code-capacity threshold for large bias where the XZZX code is effectively equivalent to the phase-correcting repetition code, with the same logical bit-flip failure rate.  


For future work it will be interesting to extend the analysis of finite size effects by studying the the statistical mechanics representations with fixed boundary conditions in a wider range of topological codes, including Clifford-deformed surface codes\cite{dua2022clifford}, and explore the generalization of the mapping to subsystem codes and phenomenological or circuit-level measurement noise~\cite{wang2003confinement,chubb2021statistical}.   

Numerical code and data related to this work is found in \href{https://github.com/yinzi-xiao/Finite-size-correction-for-surface-code}{https://github.com/yinzi-xiao/Finite-size-correction-for-surface-code}

\begin{acknowledgments}
We thank Ben Brown, Ben Criger, and Timo Hillmann for valuable discussions. We acknowledge financial support from the Knut and Alice Wallenberg Foundation through the Wallenberg Centre for Quantum Technology (WACQT). Computations were enabled by resources provided by the National Academic Infrastructure for Supercomputing in Sweden (NAISS) and the Swedish National Infrastructure for Computing (SNIC), partially funded by the Swedish Research Council through grant agreements no. 2022-06725 and no. 2018-05973, and by Chalmers e-Commons through the Chalmers Centre for Computational Science and Engineering (C3SE).

\end{acknowledgments}

\bibliography{references}

\begin{thebibliography}{10}

\bibitem{Kelly2015StateCircuit}
J.~Kelly, R.~Barends, A.~G. Fowler, A.~Megrant, E.~Jeffrey, T.~C. White,
  D.~Sank, J.~Y. Mutus, B.~Campbell, Yu~Chen, Z.~Chen, B.~Chiaro, A.~Dunsworth,
  I.-C. Hoi, C.~Neill, P.~J.~J. O’Malley, C.~Quintana, P.~Roushan,
  A.~Vainsencher, J.~Wenner, A.~N. Cleland, and John~M. Martinis.
\newblock ``{State preservation by repetitive error detection in a
  superconducting quantum circuit}''.
\newblock \href{https://dx.doi.org/10.1038/nature14270}{Nature {\bf 519},
  66--69}~(2015).

\bibitem{Takita2017ExperimentalQubits}
Maika Takita, Andrew~W. Cross, A.~D. C{\'{o}}rcoles, Jerry~M. Chow, and Jay~M.
  Gambetta.
\newblock ``{Experimental Demonstration of Fault-Tolerant State Preparation
  with Superconducting Qubits}''.
\newblock \href{https://dx.doi.org/10.1103/PhysRevLett.119.180501}{Physical
  Review Letters {\bf 119}, 180501}~(2017).

\bibitem{PhysRevA.97.052313}
James~R. Wootton and Daniel Loss.
\newblock ``Repetition code of 15 qubits''.
\newblock \href{https://dx.doi.org/10.1103/PhysRevA.97.052313}{Phys. Rev. A
  {\bf 97}, 052313}~(2018).

\bibitem{wootton2020benchmarking}
James~R Wootton.
\newblock ``Benchmarking near-term devices with quantum error correction''.
\newblock \href{https://dx.doi.org/10.1088/2058-9565/aba038}{Quantum Science
  and Technology {\bf 5}, 044004}~(2020).

\bibitem{Andersen2020RepeatedCode}
Christian~Kraglund Andersen, Ants Remm, Stefania Lazar, Sebastian Krinner,
  Nathan Lacroix, Graham~J. Norris, Mihai Gabureac, Christopher Eichler, and
  Andreas Wallraff.
\newblock ``{Repeated quantum error detection in a surface code}''.
\newblock \href{https://dx.doi.org/10.1038/s41567-020-0920-y}{Nature Physics
  {\bf 16}, 875--880}~(2020).

\bibitem{Satzinger2021Realizing}
K.~J. Satzinger et~al.
\newblock ``{Realizing topologically ordered states on a quantum processor}''.
\newblock \href{https://dx.doi.org/10.1126/science.abi8378}{Science {\bf 374},
  1237}~(2021).

\bibitem{Egan2021Fault-tolerantQubit}
Laird Egan, Dripto~M. Debroy, Crystal Noel, Andrew Risinger, Daiwei Zhu,
  Debopriyo Biswas, Michael Newman, Muyuan Li, Kenneth~R. Brown, Marko Cetina,
  and Christopher Monroe.
\newblock ``{Fault-tolerant control of an error-corrected qubit}''.
\newblock \href{https://dx.doi.org/10.1038/s41586-021-03928-y}{Nature {\bf
  598}, 281--286}~(2021).

\bibitem{Chen2021ExponentialCorrection}
Zijun Chen et~al.
\newblock ``{Exponential suppression of bit or phase errors with cyclic error
  correction}''.
\newblock \href{https://dx.doi.org/10.1038/s41586-021-03588-y}{Nature {\bf
  595}, 383--387}~(2021).

\bibitem{Erhard2021EntanglingSurgery}
Alexander Erhard, Hendrik Poulsen~Nautrup, Michael Meth, Lukas Postler, Roman
  Stricker, Martin Stadler, Vlad Negnevitsky, Martin Ringbauer, Philipp
  Schindler, Hans~J. Briegel, Rainer Blatt, Nicolai Friis, and Thomas Monz.
\newblock ``{Entangling logical qubits with lattice surgery}''.
\newblock \href{https://dx.doi.org/10.1038/s41586-020-03079-6}{Nature {\bf
  589}, 220--224}~(2021).

\bibitem{ryananderson2021realization}
C.~Ryan-Anderson, J.~G. Bohnet, K.~Lee, D.~Gresh, A.~Hankin, J.~P. Gaebler,
  D.~Francois, A.~Chernoguzov, D.~Lucchetti, N.~C. Brown, T.~M. Gatterman,
  S.~K. Halit, K.~Gilmore, J.~A. Gerber, B.~Neyenhuis, D.~Hayes, and R.~P.
  Stutz.
\newblock ``Realization of real-time fault-tolerant quantum error correction''.
\newblock \href{https://dx.doi.org/10.1103/PhysRevX.11.041058}{Phys. Rev. X
  {\bf 11}, 041058}~(2021).

\bibitem{marques2021logicalqubit}
J.~F. Marques, B.~M. Varbanov, M.~S. Moreira, H.~Ali, N.~Muthusubramanian,
  C.~Zachariadis, F.~Battistel, M.~Beekman, N.~Haider, W.~Vlothuizen, A.~Bruno,
  B.~M. Terhal, and L.~DiCarlo.
\newblock ``Logical-qubit operations in an error-detecting surface code''.
\newblock \href{https://dx.doi.org/10.1038/s41567-021-01423-9}{Nature Physics
  {\bf 18}, 80--86}~(2021).

\bibitem{Postler_2022}
Lukas Postler, Sascha Heussen, Ivan Pogorelov, Manuel Rispler, Thomas Feldker,
  Michael Meth, Christian~D. Marciniak, Roman Stricker, Martin Ringbauer,
  Rainer Blatt, Philipp Schindler, Markus Müller, and Thomas Monz.
\newblock ``Demonstration of fault-tolerant universal quantum gate
  operations''.
\newblock \href{https://dx.doi.org/10.1038/s41586-022-04721-1}{Nature {\bf
  605}, 675--680}~(2022).

\bibitem{krinner2022realizing}
Sebastian Krinner, Nathan Lacroix, Ants Remm, Agustin Di~Paolo, Elie Genois,
  Catherine Leroux, Christoph Hellings, Stefania Lazar, Christian~Kraglund
  Andersen, et~al.
\newblock ``Realizing repeated quantum error correction in a distance-three
  surface code''.
\newblock \href{https://dx.doi.org/10.1038/s41586-022-04566-8}{Nature {\bf
  605}, 669--674}~(2022).

\bibitem{Bluvstein2021}
Dolev Bluvstein, Harry Levine, Giulia Semeghini, Tout~T. Wang, Sepehr Ebadi,
  Marcin Kalinowski, Alexander Keesling, Nishad Maskara, Hannes Pichler, Markus
  Greiner, Vladan Vuleti{\'{c}}, and Mikhail~D. Lukin.
\newblock ``A quantum processor based on coherent transport of entangled atom
  arrays''.
\newblock \href{https://dx.doi.org/10.1038/s41586-022-04592-6}{Nature {\bf
  604}, 451--456}~(2022).

\bibitem{google2023suppressing}
{Google Quantum AI}.
\newblock ``Suppressing quantum errors by scaling a surface code logical
  qubit''.
\newblock \href{https://dx.doi.org/10.1038/s41586-022-05434-1}{Nature {\bf
  614}, 676--681}~(2023).

\bibitem{moses2023race}
S.~A. Moses, C.~H. Baldwin, M.~S. Allman, R.~Ancona, L.~Ascarrunz, C.~Barnes,
  J.~Bartolotta, B.~Bjork, P.~Blanchard, M.~Bohn, J.~G. Bohnet, N.~C. Brown,
  N.~Q. Burdick, W.~C. Burton, S.~L. Campbell, J.~P. Campora, C.~Carron,
  J.~Chambers, J.~W. Chan, Y.~H. Chen, A.~Chernoguzov, E.~Chertkov, J.~Colina,
  J.~P. Curtis, R.~Daniel, M.~DeCross, D.~Deen, C.~Delaney, J.~M. Dreiling,
  C.~T. Ertsgaard, J.~Esposito, B.~Estey, M.~Fabrikant, C.~Figgatt, C.~Foltz,
  M.~Foss-Feig, D.~Francois, J.~P. Gaebler, T.~M. Gatterman, C.~N. Gilbreth,
  J.~Giles, E.~Glynn, A.~Hall, A.~M. Hankin, A.~Hansen, D.~Hayes, B.~Higashi,
  I.~M. Hoffman, B.~Horning, J.~J. Hout, R.~Jacobs, J.~Johansen, L.~Jones,
  J.~Karcz, T.~Klein, P.~Lauria, P.~Lee, D.~Liefer, S.~T. Lu, D.~Lucchetti,
  C.~Lytle, A.~Malm, M.~Matheny, B.~Mathewson, K.~Mayer, D.~B. Miller,
  M.~Mills, B.~Neyenhuis, L.~Nugent, S.~Olson, J.~Parks, G.~N. Price, Z.~Price,
  M.~Pugh, A.~Ransford, A.~P. Reed, C.~Roman, M.~Rowe, C.~Ryan-Anderson,
  S.~Sanders, J.~Sedlacek, P.~Shevchuk, P.~Siegfried, T.~Skripka, B.~Spaun,
  R.~T. Sprenkle, R.~P. Stutz, M.~Swallows, R.~I. Tobey, A.~Tran, T.~Tran,
  E.~Vogt, C.~Volin, J.~Walker, A.~M. Zolot, and J.~M. Pino.
\newblock ``A race-track trapped-ion quantum processor''.
\newblock \href{https://dx.doi.org/10.1103/PhysRevX.13.041052}{Phys. Rev. X
  {\bf 13}, 041052}~(2023).

\bibitem{sundaresan2023demonstrating}
Neereja Sundaresan, Theodore~J Yoder, Youngseok Kim, Muyuan Li, Edward~H Chen,
  Grace Harper, Ted Thorbeck, Andrew~W Cross, Antonio~D C{\'o}rcoles, and Maika
  Takita.
\newblock ``Demonstrating multi-round subsystem quantum error correction using
  matching and maximum likelihood decoders''.
\newblock \href{https://dx.doi.org/10.1038/s41467-023-38247-5}{Nature
  Communications {\bf 14}, 2852}~(2023).

\bibitem{Gupta2023Encoding}
Riddhi~S Gupta, Neereja Sundaresan, Thomas Alexander, Christopher~J Wood,
  Seth~T Merkel, Michael~B Healy, Marius Hillenbrand, Tomas Jochym-O’Connor,
  James~R Wootton, Theodore~J Yoder, et~al.
\newblock ``Encoding a magic state with beyond break-even fidelity''.
\newblock \href{https://dx.doi.org/10.1038/s41586-023-06846-3}{Nature {\bf
  625}, 259--263}~(2024).

\bibitem{Brown2023Advances}
Natalie~C. Brown, John Peter Campora~III au2, Cassandra Granade, Bettina Heim,
  Stefan Wernli, Ciaran Ryan-Anderson, Dominic Lucchetti, Adam Paetznick,
  Martin Roetteler, Krysta Svore, and Alex Chernoguzov.
\newblock ``Advances in compilation for quantum hardware -- a demonstration of
  magic state distillation and repeat-until-success protocols''~(2023).
\newblock  \href{http://arxiv.org/abs/2310.12106}{arXiv:2310.12106}.

\bibitem{Wang2023FT}
Yang Wang, Selwyn Simsek, Thomas~M Gatterman, Justin~A Gerber, Kevin Gilmore,
  Dan Gresh, Nathan Hewitt, Chandler~V Horst, Mitchell Matheny, Tanner Mengle,
  et~al.
\newblock ``Fault-tolerant one-bit addition with the smallest interesting color
  code''.
\newblock \href{https://dx.doi.org/10.1126/sciadv.ado9024}{Science Advances
  {\bf 10}, eado9024}~(2024).

\bibitem{Bluvstein2023}
Dolev Bluvstein, Simon~J Evered, Alexandra~A Geim, Sophie~H Li, Hengyun Zhou,
  Tom Manovitz, Sepehr Ebadi, Madelyn Cain, Marcin Kalinowski, Dominik
  Hangleiter, et~al.
\newblock ``Logical quantum processor based on reconfigurable atom arrays''.
\newblock \href{https://dx.doi.org/10.1038/s41586-023-06927-3}{Nature {\bf
  626}, 58--65}~(2024).

\bibitem{PhysRevA.32.3266}
Asher Peres.
\newblock ``Reversible logic and quantum computers''.
\newblock \href{https://dx.doi.org/10.1103/PhysRevA.32.3266}{Phys. Rev. A {\bf
  32}, 3266--3276}~(1985).

\bibitem{PhysRevA.52.R2493}
Peter~W. Shor.
\newblock ``Scheme for reducing decoherence in quantum computer memory''.
\newblock \href{https://dx.doi.org/10.1103/PhysRevA.52.R2493}{Phys. Rev. A {\bf
  52}, R2493--R2496}~(1995).

\bibitem{RevModPhys.87.307}
Barbara~M. Terhal.
\newblock ``Quantum error correction for quantum memories''.
\newblock \href{https://dx.doi.org/10.1103/RevModPhys.87.307}{Rev. Mod. Phys.
  {\bf 87}, 307--346}~(2015).

\bibitem{Girvin_2023}
Steven~M. Girvin.
\newblock ``Introduction to quantum error correction and fault tolerance''.
\newblock \href{https://dx.doi.org/10.21468/scipostphyslectnotes.70}{SciPost
  Physics Lecture Notes}~(2023).

\bibitem{Kitaev2003Fault-tolerantAnyons}
A.Yu. Kitaev.
\newblock ``{Fault-tolerant quantum computation by anyons}''.
\newblock \href{https://dx.doi.org/10.1016/S0003-4916(02)00018-0}{Annals of
  Physics {\bf 303}, 2--30}~(2003).

\bibitem{dennis2002topological}
Eric Dennis, Alexei Kitaev, Andrew Landahl, and John Preskill.
\newblock ``Topological quantum memory''.
\newblock \href{https://dx.doi.org/10.1063/1.1499754}{Journal of Mathematical
  Physics {\bf 43}, 4452--4505}~(2002).

\bibitem{PhysRevLett.98.190504}
Robert Raussendorf and Jim Harrington.
\newblock ``Fault-tolerant quantum computation with high threshold in two
  dimensions''.
\newblock \href{https://dx.doi.org/10.1103/PhysRevLett.98.190504}{Phys. Rev.
  Lett. {\bf 98}, 190504}~(2007).

\bibitem{PhysRevA.86.032324}
Austin~G. Fowler, Matteo Mariantoni, John~M. Martinis, and Andrew~N. Cleland.
\newblock ``Surface codes: Towards practical large-scale quantum computation''.
\newblock \href{https://dx.doi.org/10.1103/PhysRevA.86.032324}{Phys. Rev. A
  {\bf 86}, 032324}~(2012).

\bibitem{Bombin2007OptimalStudy}
H.~Bombin and M.~A. Martin-Delgado.
\newblock ``{Optimal resources for topological two-dimensional stabilizer
  codes: Comparative study}''.
\newblock \href{https://dx.doi.org/10.1103/PhysRevA.76.012305}{Physical Review
  A {\bf 76}, 012305}~(2007).

\bibitem{Chamberland2020}
Christopher Chamberland, Guanyu Zhu, Theodore~J. Yoder, Jared~B. Hertzberg, and
  Andrew~W. Cross.
\newblock ``Topological and subsystem codes on low-degree graphs with flag
  qubits''.
\newblock \href{https://dx.doi.org/10.1103/physrevx.10.011022}{Physical Review
  X{\bf 10}}~(2020).

\bibitem{Hastings2021DGLT}
Matthew~B. Hastings and Jeongwan Haah.
\newblock ``{Dynamically Generated Logical Qubits}''.
\newblock \href{https://dx.doi.org/10.22331/q-2021-10-19-564}{Quantum {\bf 5},
  564}~(2021).

\bibitem{Wootton2015}
James~R Wootton.
\newblock ``A family of stabilizer codes for {$D(Z_2)$} anyons and majorana
  modes''.
\newblock \href{https://dx.doi.org/10.1088/1751-8113/48/21/215302}{Journal of
  Physics A: Mathematical and Theoretical {\bf 48}, 215302}~(2015).

\bibitem{Wootton2021}
James~R. Wootton.
\newblock ``{Hexagonal matching codes with 2-body measurements}''~(2021).
\newblock  \href{http://arxiv.org/abs/2109.13308}{arXiv:2109.13308}.

\bibitem{PhysRevLett.124.130501}
David~K. Tuckett, Stephen~D. Bartlett, Steven~T. Flammia, and Benjamin~J.
  Brown.
\newblock ``Fault-tolerant thresholds for the surface code in excess of 5\%
  under biased noise''.
\newblock \href{https://dx.doi.org/10.1103/PhysRevLett.124.130501}{Phys. Rev.
  Lett. {\bf 124}, 130501}~(2020).

\bibitem{PhysRevX.9.041031}
David~K. Tuckett, Andrew~S. Darmawan, Christopher~T. Chubb, Sergey Bravyi,
  Stephen~D. Bartlett, and Steven~T. Flammia.
\newblock ``Tailoring surface codes for highly biased noise''.
\newblock \href{https://dx.doi.org/10.1103/PhysRevX.9.041031}{Phys. Rev. X {\bf
  9}, 041031}~(2019).

\bibitem{Ataides2021XZZX}
J.~Pablo Bonilla~Ataides, David~K. Tuckett, Stephen~D. Bartlett, Steven~T.
  Flammia, and Benjamin~J. Brown.
\newblock ``{The XZZX surface code}''.
\newblock \href{https://dx.doi.org/10.1038/s41467-021-22274-1}{Nature
  Communications {\bf 12}, 2172}~(2021).

\bibitem{Srivastava2022xyzhexagonal}
Basudha Srivastava, Anton Frisk~Kockum, and Mats Granath.
\newblock ``The {XYZ}{$^2$} hexagonal stabilizer code''.
\newblock \href{https://dx.doi.org/10.22331/q-2022-04-27-698}{{Quantum} {\bf
  6}, 698}~(2022).

\bibitem{PRXQuantum.2.030345}
Andrew~S. Darmawan, Benjamin~J. Brown, Arne~L. Grimsmo, David~K. Tuckett, and
  Shruti Puri.
\newblock ``Practical quantum error correction with the xzzx code and kerr-cat
  qubits''.
\newblock \href{https://dx.doi.org/10.1103/PRXQuantum.2.030345}{PRX Quantum
  {\bf 2}, 030345}~(2021).

\bibitem{tiurev2023correcting}
Konstantin Tiurev, Peter-Jan H.~S. Derks, Joschka Roffe, Jens Eisert, and
  Jan-Michael Reiner.
\newblock ``Correcting non-independent and non-identically distributed errors
  with surface codes''.
\newblock \href{https://dx.doi.org/10.22331/q-2023-09-26-1123}{{Quantum} {\bf
  7}, 1123}~(2023).

\bibitem{Huang2022}
Eric Huang, Arthur Pesah, Christopher~T. Chubb, Michael Vasmer, and Arpit Dua.
\newblock ``Tailoring three-dimensional topological codes for biased noise''.
\newblock \href{https://dx.doi.org/10.1103/PRXQuantum.4.030338}{PRX Quantum
  {\bf 4}, 030338}~(2023).

\bibitem{Tiurev2023Domain}
Konstantin Tiurev, Arthur Pesah, Peter-Jan H.~S. Derks, Joschka Roffe, Jens
  Eisert, Markus~S. Kesselring, and Jan-Michael Reiner.
\newblock ``The domain wall color code''~(2024).
\newblock  \href{http://arxiv.org/abs/2307.00054}{arXiv:2307.00054}.

\bibitem{gottesman2014faulttolerant}
Daniel Gottesman.
\newblock ``Fault-tolerant quantum computation with constant overhead''~(2014).
\newblock  \href{http://arxiv.org/abs/1310.2984}{arXiv:1310.2984}.

\bibitem{PRXQuantum.2.040101}
Nikolas~P. Breuckmann and Jens~Niklas Eberhardt.
\newblock ``Quantum low-density parity-check codes''.
\newblock \href{https://dx.doi.org/10.1103/PRXQuantum.2.040101}{PRX Quantum
  {\bf 2}, 040101}~(2021).

\bibitem{bravyi2023highthreshold}
Sergey Bravyi, Andrew~W Cross, Jay~M Gambetta, Dmitri Maslov, Patrick Rall, and
  Theodore~J Yoder.
\newblock ``High-threshold and low-overhead fault-tolerant quantum memory''.
\newblock \href{https://dx.doi.org/10.1038/s41586-024-07107-7}{Nature {\bf
  627}, 778--782}~(2024).

\bibitem{Tuckett2018UltrahighNoise}
David~K. Tuckett, Stephen~D. Bartlett, and Steven~T. Flammia.
\newblock ``{Ultrahigh Error Threshold for Surface Codes with Biased Noise}''.
\newblock \href{https://dx.doi.org/10.1103/PHYSREVLETT.120.050505}{Physical
  Review Letters {\bf 120}, 050505}~(2018).

\bibitem{dua2022clifford}
Arpit Dua, Aleksander Kubica, Liang Jiang, Steven~T. Flammia, and Michael~J.
  Gullans.
\newblock ``Clifford-deformed surface codes''.
\newblock \href{https://dx.doi.org/10.1103/PRXQuantum.5.010347}{PRX Quantum
  {\bf 5}, 010347}~(2024).

\bibitem{PhysRevA.54.3824}
Charles~H. Bennett, David~P. DiVincenzo, John~A. Smolin, and William~K.
  Wootters.
\newblock ``Mixed-state entanglement and quantum error correction''.
\newblock \href{https://dx.doi.org/10.1103/PhysRevA.54.3824}{Phys. Rev. A {\bf
  54}, 3824--3851}~(1996).

\bibitem{wilde2011classical}
Mark~M Wilde.
\newblock ``From classical to quantum shannon theory''~(2011).
\newblock
  url:~\href{https://doi.org/10.48550/arXiv.1106.1445}{doi.org/10.48550/arXiv.1106.1445}.

\bibitem{wang2003confinement}
Chenyang Wang, Jim Harrington, and John Preskill.
\newblock ``Confinement-higgs transition in a disordered gauge theory and the
  accuracy threshold for quantum memory''.
\newblock \href{https://dx.doi.org/10.1016/S0003-4916(02)00019-2}{Annals of
  Physics {\bf 303}, 31--58}~(2003).

\bibitem{Wootton2012HighCode}
James~R. Wootton and Daniel Loss.
\newblock ``{High Threshold Error Correction for the Surface Code}''.
\newblock \href{https://dx.doi.org/10.1103/PhysRevLett.109.160503}{Physical
  Review Letters {\bf 109}, 160503}~(2012).

\bibitem{Hutter2014EfficientCode}
Adrian Hutter, James~R. Wootton, and Daniel Loss.
\newblock ``{Efficient Markov chain Monte Carlo algorithm for the surface
  code}''.
\newblock \href{https://dx.doi.org/10.1103/PhysRevA.89.022326}{Physical Review
  A {\bf 89}, 022326}~(2014).

\bibitem{Bravyi2014EfficientCode}
Sergey Bravyi, Martin Suchara, and Alexander Vargo.
\newblock ``{Efficient algorithms for maximum likelihood decoding in the
  surface code}''.
\newblock \href{https://dx.doi.org/10.1103/PhysRevA.90.032326}{Physical Review
  A {\bf 90}, 032326}~(2014).

\bibitem{Hammar_2022}
Karl Hammar, Alexei Orekhov, Patrik~Wallin Hybelius, Anna~Katariina Wisakanto,
  Basudha Srivastava, Anton~Frisk Kockum, and Mats Granath.
\newblock ``Error-rate-agnostic decoding of topological stabilizer codes''.
\newblock \href{https://dx.doi.org/10.1103/PhysRevA.105.042616}{Phys. Rev. A
  {\bf 105}, 042616}~(2022).

\bibitem{qecsim}
David~Kingsley Tuckett.
\newblock ``Tailoring surface codes: Improvements in quantum error correction
  with biased noise''.
\newblock \href{https://dx.doi.org/10.25910/x8xw-9077}{PhD thesis}.
\newblock University of Sydney.
\newblock ~(2020).

\bibitem{Chubb2021TensorNetwork}
Christopher~T Chubb.
\newblock ``General tensor network decoding of 2d pauli codes''~(2021).
\newblock
  url:~\href{https://arxiv.org/abs/2101.04125}{arxiv.org/abs/2101.04125}.

\bibitem{lange2023datadriven}
Moritz Lange, Pontus Havström, Basudha Srivastava, Valdemar Bergentall, Karl
  Hammar, Olivia Heuts, Evert van Nieuwenburg, and Mats Granath.
\newblock ``Data-driven decoding of quantum error correcting codes using graph
  neural networks''~(2023).
\newblock  \href{http://arxiv.org/abs/2307.01241}{arXiv:2307.01241}.

\bibitem{varbanov2023neural}
Boris~M. Varbanov, Marc Serra-Peralta, David Byfield, and Barbara~M. Terhal.
\newblock ``Neural network decoder for near-term surface-code
  experiments''~(2023).
\newblock  \href{http://arxiv.org/abs/2307.03280}{arXiv:2307.03280}.

\bibitem{bausch2023learning}
Johannes Bausch, Andrew~W Senior, Francisco J~H Heras, Thomas Edlich, Alex
  Davies, Michael Newman, Cody Jones, Kevin Satzinger, Murphy~Yuezhen Niu, Sam
  Blackwell, George Holland, Dvir Kafri, Juan Atalaya, Craig Gidney, Demis
  Hassabis, Sergio Boixo, Hartmut Neven, and Pushmeet Kohli.
\newblock ``Learning to decode the surface code with a recurrent,
  transformer-based neural network''~(2023).
\newblock  \href{http://arxiv.org/abs/2310.05900}{arXiv:2310.05900}.

\bibitem{gidney2021stim}
Craig Gidney.
\newblock ``Stim: a fast stabilizer circuit simulator''.
\newblock \href{https://dx.doi.org/10.22331/q-2021-07-06-497}{{Quantum} {\bf
  5}, 497}~(2021).

\bibitem{PhysRevA.108.022401}
Antonio~deMarti iOlius, Josu~Etxezarreta Martinez, Patricio Fuentes, and
  Pedro~M. Crespo.
\newblock ``Performance enhancement of surface codes via recursive
  minimum-weight perfect-match decoding''.
\newblock \href{https://dx.doi.org/10.1103/PhysRevA.108.022401}{Phys. Rev. A
  {\bf 108}, 022401}~(2023).

\bibitem{PhysRevX.2.021004}
H.~Bombin, Ruben~S. Andrist, Masayuki Ohzeki, Helmut~G. Katzgraber, and M.~A.
  Martin-Delgado.
\newblock ``Strong resilience of topological codes to depolarization''.
\newblock \href{https://dx.doi.org/10.1103/PhysRevX.2.021004}{Phys. Rev. X {\bf
  2}, 021004}~(2012).

\bibitem{kovalev2014spin}
Alexey~A. Kovalev and Leonid~P. Pryadko.
\newblock ``Spin glass reflection of the decoding transition for quantum error
  correcting codes''~(2014).
\newblock  \href{http://arxiv.org/abs/1311.7688}{arXiv:1311.7688}.

\bibitem{chubb2021statistical}
Christopher~T Chubb and Steven~T Flammia.
\newblock ``Statistical mechanical models for quantum codes with correlated
  noise''.
\newblock \href{https://dx.doi.org/10.4171/aihpd/105}{Annales de l’Institut
  Henri Poincar{\'e} D {\bf 8}, 269--321}~(2021).

\bibitem{Vodola2022fundamental}
Davide Vodola, Manuel Rispler, Seyong Kim, and Markus M{\"{u}}ller.
\newblock ``Fundamental thresholds of realistic quantum error correction
  circuits from classical spin models''.
\newblock \href{https://dx.doi.org/10.22331/q-2022-01-05-618}{{Quantum} {\bf
  6}, 618}~(2022).

\bibitem{PhysRevLett.131.060603}
Florian Venn, Jan Behrends, and Benjamin B\'eri.
\newblock ``Coherent-error threshold for surface codes from majorana
  delocalization''.
\newblock \href{https://dx.doi.org/10.1103/PhysRevLett.131.060603}{Phys. Rev.
  Lett. {\bf 131}, 060603}~(2023).

\bibitem{nishimori1981internal}
Hidetoshi Nishimori.
\newblock ``Internal energy, specific heat and correlation function of the
  bond-random ising model''.
\newblock \href{https://dx.doi.org/10.1143/PTP.66.1169}{Progress of Theoretical
  Physics {\bf 66}, 1169--1181}~(1981).

\bibitem{sutherland1970two}
Bill Sutherland.
\newblock ``Two-dimensional hydrogen bonded crystals without the ice rule''.
\newblock \href{https://dx.doi.org/10.1063/1.1665111}{Journal of Mathematical
  Physics {\bf 11}, 3183--3186}~(1970).

\bibitem{baxter1971eight}
R.~J. Baxter.
\newblock ``Eight-vertex model in lattice statistics''.
\newblock \href{https://dx.doi.org/10.1103/PhysRevLett.26.832}{Phys. Rev. Lett.
  {\bf 26}, 832--833}~(1971).

\bibitem{PhysRevB.12.429}
C.~S. Hsue, K.~Y. Lin, and F.~Y. Wu.
\newblock ``Staggered eight-vertex model''.
\newblock \href{https://dx.doi.org/10.1103/PhysRevB.12.429}{Phys. Rev. B {\bf
  12}, 429--437}~(1975).

\bibitem{fan1972critical}
C~Fan.
\newblock ``On critical properties of the ashkin-teller model''.
\newblock \href{https://dx.doi.org/10.1016/0375-9601(72)91051-1}{Physics
  Letters A {\bf 39}, 136}~(1972).

\bibitem{PhysRev.64.178}
J.~Ashkin and E.~Teller.
\newblock ``Statistics of two-dimensional lattices with four components''.
\newblock \href{https://dx.doi.org/10.1103/PhysRev.64.178}{Phys. Rev. {\bf 64},
  178--184}~(1943).

\bibitem{wu1971ising}
F.~W. Wu.
\newblock ``Ising model with four-spin interactions''.
\newblock \href{https://dx.doi.org/10.1103/PhysRevB.4.2312}{Phys. Rev. B {\bf
  4}, 2312--2314}~(1971).

\bibitem{baxter2016exactly}
Rodney~J Baxter.
\newblock ``Exactly solved models in statistical mechanics''.
\newblock \href{https://dx.doi.org/10.1142/9789814415255_0002}{Elsevier}.
  ~(1982).

\bibitem{harrington2004analysis}
James~William Harrington.
\newblock ``Analysis of quantum error-correcting codes: symplectic lattice
  codes and toric codes''.
\newblock \href{https://dx.doi.org/10.7907/AHMQ-EG82}{California Institute of
  Technology}. ~(2004).

\bibitem{goldenfeld2018lectures}
Nigel Goldenfeld.
\newblock ``Lectures on phase transitions and the renormalization group''.
\newblock CRC Press. ~(2018).
\newblock
  url:~\href{https://doi.org/10.1201/9780429493492}{doi.org/10.1201/9780429493492}.

\bibitem{PhysRevX.9.041053}
J\'er\'emie Guillaud and Mazyar Mirrahimi.
\newblock ``Repetition cat qubits for fault-tolerant quantum computation''.
\newblock \href{https://dx.doi.org/10.1103/PhysRevX.9.041053}{Phys. Rev. X {\bf
  9}, 041053}~(2019).

\bibitem{PRXQuantum.3.010329}
Christopher Chamberland, Kyungjoo Noh, Patricio Arrangoiz-Arriola, Earl~T.
  Campbell, Connor~T. Hann, Joseph Iverson, Harald Putterman, Thomas~C.
  Bohdanowicz, Steven~T. Flammia, Andrew Keller, Gil Refael, John Preskill,
  Liang Jiang, Amir~H. Safavi-Naeini, Oskar Painter, and Fernando~G.S.L.
  Brand\~ao.
\newblock ``Building a fault-tolerant quantum computer using concatenated cat
  codes''.
\newblock \href{https://dx.doi.org/10.1103/PRXQuantum.3.010329}{PRX Quantum
  {\bf 3}, 010329}~(2022).

\bibitem{PhysRevX.13.031007}
Oscar Higgott, Thomas~C. Bohdanowicz, Aleksander Kubica, Steven~T. Flammia, and
  Earl~T. Campbell.
\newblock ``Improved decoding of circuit noise and fragile boundaries of
  tailored surface codes''.
\newblock \href{https://dx.doi.org/10.1103/PhysRevX.13.031007}{Phys. Rev. X
  {\bf 13}, 031007}~(2023).

\bibitem{PhysRevB.65.054425}
F.~Merz and J.~T. Chalker.
\newblock ``Two-dimensional random-bond ising model, free fermions, and the
  network model''.
\newblock \href{https://dx.doi.org/10.1103/PhysRevB.65.054425}{Phys. Rev. B
  {\bf 65}, 054425}~(2002).

\bibitem{PhysRevLett.87.047201}
A.~Honecker, M.~Picco, and P.~Pujol.
\newblock ``{Universality Class of the Nishimori Point in the 2D
  $\ifmmode\pm\else\textpm\fi{}\mathit{J}$ Random-Bond Ising Model}''.
\newblock \href{https://dx.doi.org/10.1103/PhysRevLett.87.047201}{Phys. Rev.
  Lett. {\bf 87}, 047201}~(2001).

\end{thebibliography}

\appendix

\section{Hashing bound for random stabilizer code}
\label{app:hashing}
Here we reformulate and outline the proof of the hashing bound for the threshold of a random stabilizer code under i.i.d. Pauli noise, following~\cite{wilde2011classical}.  
For a stabilizer code over $n$ qubits with a single logical qubit, there are $N_P=2^{n-1}$ sectors of the Hilbert space, specified by distinct syndromes. The probability of an arbitrary error chain is given by $\pi_C=\prod_i(p_i)^{n_i}=2^{\sum_i n_i\lg_2p_i}$, where $p_i\in (p_I,p_x,p_y,p_z)$, with $p_I=1-p$ and $n_i$ is the number of elements of each type (including $I$) in $C$. In the limit of large $n$ the distribution of error chains will be sharply peaked at the mean, $n_i\approx \langle n_i\rangle=np_i$. The probability of such a typical chain is given by 
$\pi_{C_T}=2^{-nH(p)}$, where $H(p)=-\sum_ip_i\log_2p_i$ is the entropy of the error channel. The typical chains make up the bulk of the probability mass, such that the number of chains is given $N_T=2^{nH(p)}$. (Equivalently, this corresponds to $N_T\approx \binom{n}{np_x,np_y,np_x,np_I}$, the number of chains with fixed numbers of respective errors.) Given a random stabilizer code, such that the check operators are not correlated with the error model, we can assume that the typical error chains are evenly distributed over the syndromes. The failure rate is then given by $P_f\lesssim \sum_{C_T}\pi_{C_T}(\frac{N_T}{N_P})\approx \frac{N_T}{N_P}$, where $\frac{N_T}{N_P}\approx 2^{-n(1-H(p))}$ is the probability that a second error chain is found in the same sector as $C_T$, which is a necessary requirement for failure. (In fact, an overestimate of the failure rate, as it ignores the fact that two chains in the same equivalence class will not cause a failure.) For $H(p)<1$ the failure rate will go to zero for large $n$, which gives the threshold error rate $p_c:H(p_c)=1$. 

In summary, the basic motivation is that for low error rates the number of error chains that are likely to occur are much fewer than the number of possible syndromes. Unless the code is structured in a way that gives a degeneracy of these chains, error correction will succeed. Note that the result does not rule out that a particular code  has a higher threshold. 

\section{One-dimensional $\pm K$ Ising model with open or fixed boundary conditions}
\label{app:one-d-ising}
For completeness we review the calculation of the partition function for the one-dimensional Ising model, depending on boundary conditions (see, e.g.\ \cite{goldenfeld2018lectures}). The Hamiltonian over $n$ interactions, with $n+1$ spins is given by $H=\sum_{j=1}^nK_js_{j-1}s_{j}$, with the partition function $\mathcal{Z}_{\text{1D}}=\sum_{\{s\}=\pm 1}e^{-H(\{s\})}$. If there are fixed boundary spins at either or both edges these should not be summed over. We're interested in comparing models that only differ by sign changes of one or more of the couplings, corresponding to the different equivalence classes. 

First, if neither, or only one, of the end spins are fixed the partition function is invariant with respect to sign changes of any number of couplings. This is clear from the fact that these can be absorbed by redefining the sign of subsequent spins. 

The partition function $\mathcal{Z}_{\text{1D,fixed}}$, with both edges fixed, i.e.\ $s_0=s_n=+1$, remains to be analyzed.  As the Hamiltonian is invariant under a global change $s_j\rightarrow -s_j, \forall j$, we find $\mathcal{Z}_{\text{1D,fixed}}=\frac{1}{2}\mathcal{Z}_{\text{1D,pbc}}$, where the latter is the partition function with periodic boundary conditions $s_0=s_n=\pm 1$. To solve this we can represent the partition function as a matrix product, using the transfer matrix $T^j=\left (\begin{smallmatrix} e^{-K_j} & e^{K_j} \\ e^{K_j} & e^{-K_j} \end{smallmatrix}\right )$, corresponding to $\pm 1$ entries of the spins coupled to $K_j$. The transfer matrix has  eigenvectors $(1,\pm 1)$, with eigenvalues $\lambda^j_\pm=2\cosh(-K_j),2\sinh(-K_j)$. The partition function is thus given by $\mathcal{Z}_{\text{1D,pbc}}=\text{Tr}(\Pi_j T^j)=\text{Tr}(\Pi_jU^\dagger\Lambda^jU)=\Pi_j\lambda^j_++\Pi_j\lambda^j_-$, where $U=\left (\begin{smallmatrix} 1 & 1 \\ 1 & -1 \end{smallmatrix}\right )/\sqrt{2}$ and $\Lambda^j=\left (\begin{smallmatrix} \lambda^j_+ & 0 \\ 0 & \lambda^j_- \end{smallmatrix}\right )$. 

Assuming, as relevant here, $K_j=\tau_j K$, with $\tau_j=\pm 1$, we find 
\begin{equation}
    \mathcal{Z}_{\text{1D,pbc}}=2^n\cosh^n(K)(1+\tau_t\tanh^n(-K))\,
\end{equation} 
where $\tau_t=\Pi_j\tau_j=\pm 1$ depends only on the parity of the number of flipped signs. 
For $Z$-biased noise at the special point $p_z=1-p$, and with $K=K_z$, we find $\tanh(-K)=\frac{1-1/2\eta}{1+1/2\eta}$, which gives the relative probabilities of the equivalence classes and the corresponding logical failure rates. Explicitly, Eqn.\ \ref{eq:class_prob_Z} follows from $\frac{P_\mathcal{X}}{P_\mathcal{I}}=\frac{\mathcal{Z}_{\text{1D,pbc}}(\tau_t=-1)}{\mathcal{Z}_{\text{1D,pbc}}(\tau_t=+1)}$, for $n=d_Z$, using the identity $\text{artanh}(x)=\frac{1}{2}\ln(\frac{1+x}{1-x})$ with $x=-1/2\eta$, and some elementary algebra.

\section{The XZ surface code for $K_y=0$}
\label{app:xz_surface} 
The standard surface code with X and Z stabilizers simplifies for $K_y=0$, where it becomes two decoupled random bond Ising models. This can be seen from considering, equivalently, the XY code, Eqn.\ \ref{Eq:XY_H}, for $K_z=0$. The constraint $K_y=0$ can be conveniently described in terms of two independent error rates $p_1$ and $p_2$, such that $p_x=p_1(1-p_2)$, $p_z=p_2(1-p_1)$, and $p_y=p_1p_2$.  Sublattice $A$ and $B$ have coupling constants $K_z=-\frac{1}{2}\ln\frac{1-p_1}{p_1}$ and $K_x=-\frac{1}{2}\ln\frac{1-p_2}{p_2}$, respectively, with the corresponding quenched disorder rate $p_1$ and $p_2$. The logical operator $Z_L/X_L$ couple exclusively to sublattice $B/A$, such that logical bit- and phase-flip errors are decoupled. For thresholds we can refer to the well established results for the threshold $p_{c,b}\approx 0.109$~\cite{wang2003confinement,PhysRevB.65.054425,PhysRevLett.87.047201}  for bit-flip errors on the surface code. In terms of decoding, this also means that decoding $X$ and $Z$ stabilizers separately is optimal. 

Focusing on $Z$-biased noise the point that corresponds to such decoupling is the special point $p_z=1-p$, where both $K_y=0$ and $K_x=0$. Here $p_2=0.5$ and $p_1=\frac{1}{1+2\eta}$. Whereas the model is disordered to logical phase-flip errors, we find that for $\eta$ such that $\frac{1}{1+2\eta}<p_{c,b}$, i.e. $\eta\gtrsim 4$, the threshold for logical bit-flip errors is above the point $p=p_s>0.5$. The latter implies that even for such quite small bias, logical bit-flip errors are exponentially suppressed with code distance for any relevant error rate $p<0.5$. 

Away from the special point, $K_y\neq 0$, the two sublattices are coupled and decoding the two sublattices independently will not be optimal. Nevertheless, the two decoupled $Z_2$ symmetries (flipping the signs of all spins on sublattice $A$ or $B$ independently) implies that the model has two thresholds. An interesting avenue to explore the implications of separately decoding the two sublattices, with details left for future work, is to make a mean-field decoupling. Knowing that the A sublattice has a higher threshold one can assume that sublattice B is disordered, $\langle s_Bs_B'\rangle\approx 0$, such that the four-spin term $K_y$ can be ignored, giving a RBIM with coupling $K_A=K_z$. Subsequently, deep in the ordered phase for the A sublattice, we can solve the RBIM for the B sublattice with $K_ys_As_A's_Bs_B'\approx K_y\langle s_As_A'\rangle s_Bs_B' \approx K_ys_Bs_B'$, such that the total coupling constant is $K_B\approx K_x+K_y$.   
Interestingly, except at the special point, the RBIMs for the two mean-field decoupled models are not on the Nishimori line, i.e.\ the thermal and disorder temperatures do not coincide. 
 

\section{Numerical threshold fits}
\label{app:thresholds}

\begin{figure*}\centering
\subfloat[]{\includegraphics[width=0.5\linewidth, trim={0.4cm 0.2cm 1.6cm 1cm},clip]{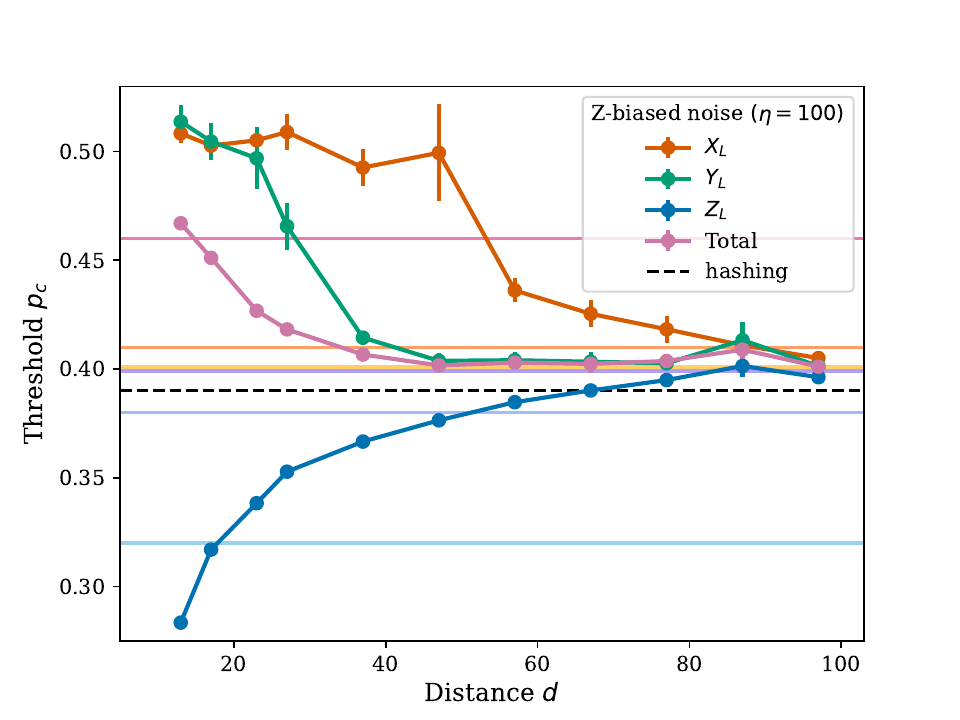}}\par
\subfloat[]{\includegraphics[width=0.5\linewidth, trim={0.4cm 0.2cm 1.2cm 1cm},clip]{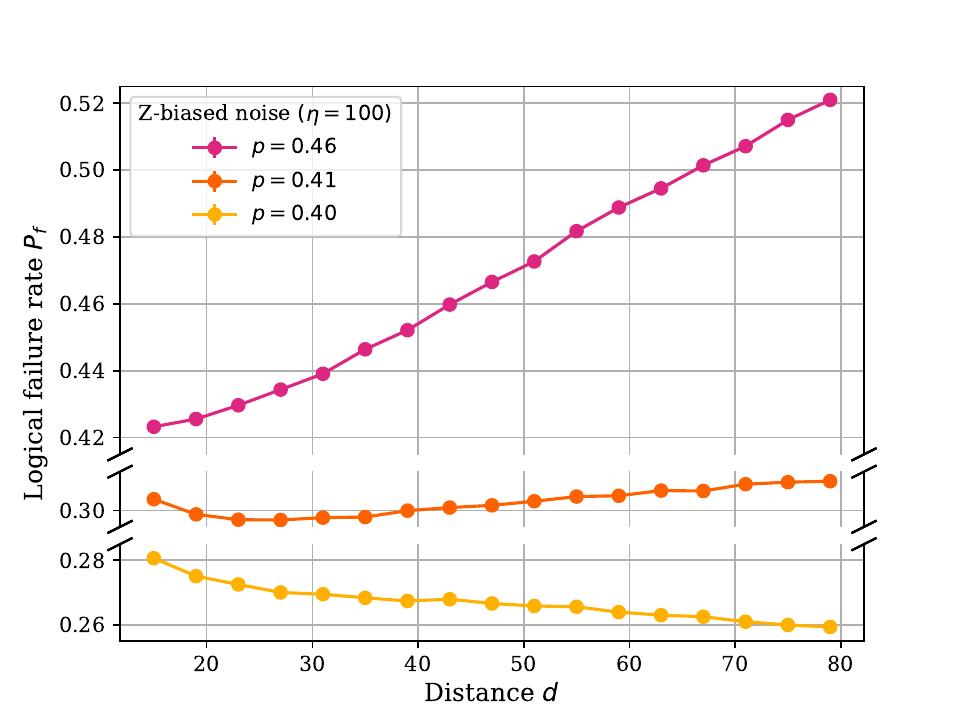}}\hfill
\subfloat[]{\includegraphics[width=0.5\linewidth, trim={0.4cm 0.2cm 1.2cm 1cm},clip]{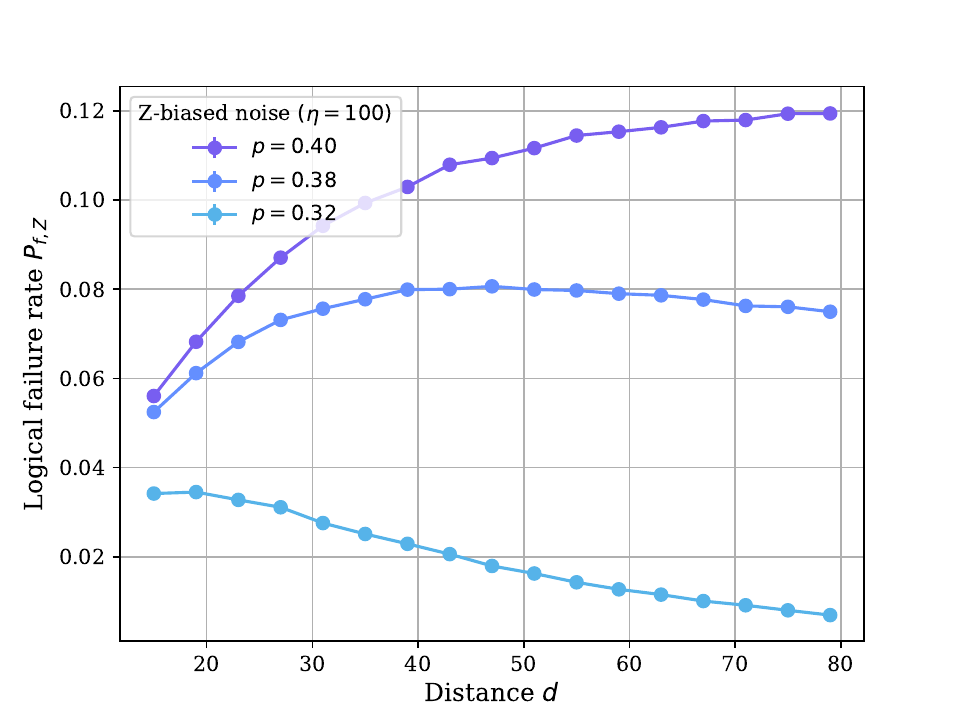}}
\caption{Estimated threshold values versus code-distance $d$, for the XZZX code, as in Fig. \ref{fig:thresholds}, at bias $\eta=100$. Thresholds converge to a value above the hashing bound (a). Constant error rate cuts for failure rates versus code-distance, showing non-monotonous behavior due to finite size corrections. Each data-point averaged over 600,000 syndromes (b, c).}
\label{fig:eta100}
\end{figure*}

\begin{figure*}\centering
\subfloat[]{\includegraphics[width=0.5\linewidth, trim={0.4cm 0.2cm 1.6cm 1cm},clip]{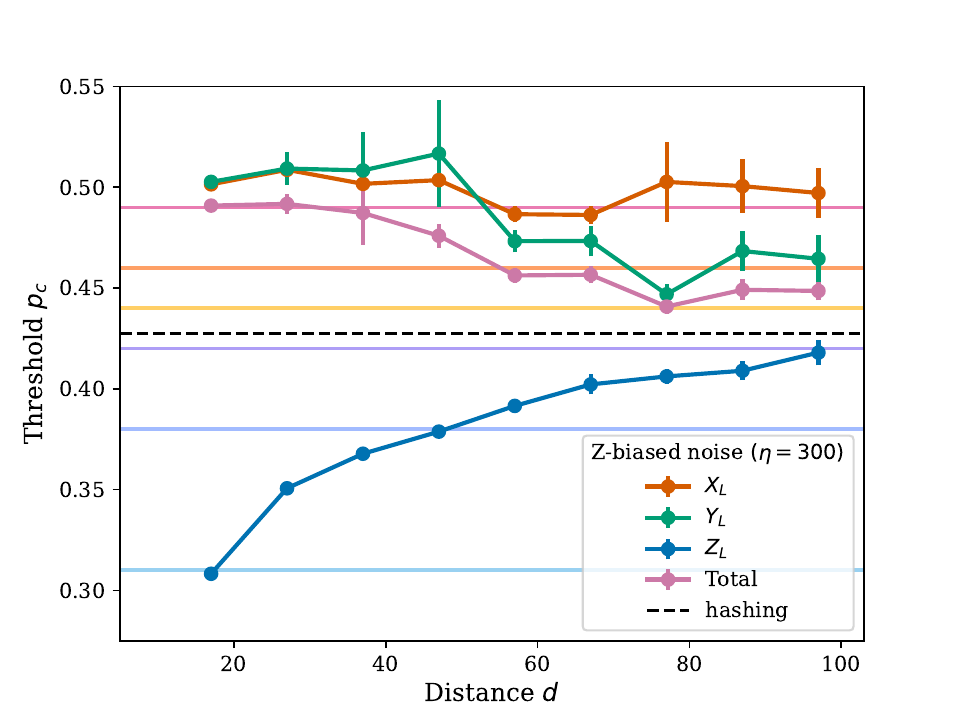}}\par
\subfloat[]{\includegraphics[width=0.5\linewidth, trim={0.4cm 0.2cm 1.2cm 1cm},clip]{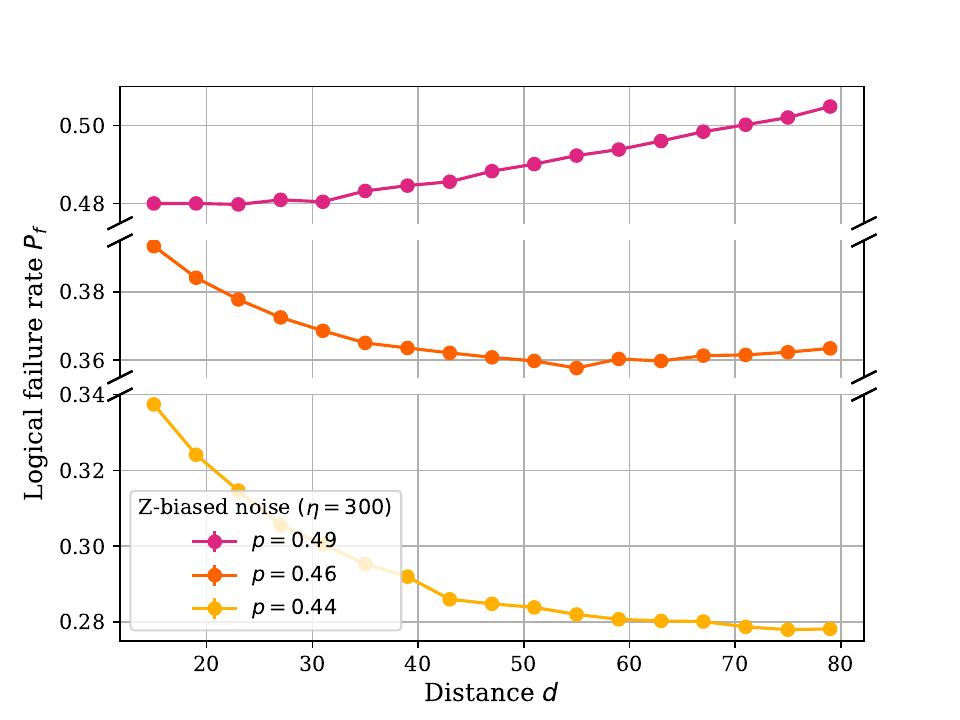}}\hfill
\subfloat[]{\includegraphics[width=0.5\linewidth, trim={0.4cm 0.2cm 1.2cm 1cm},clip]{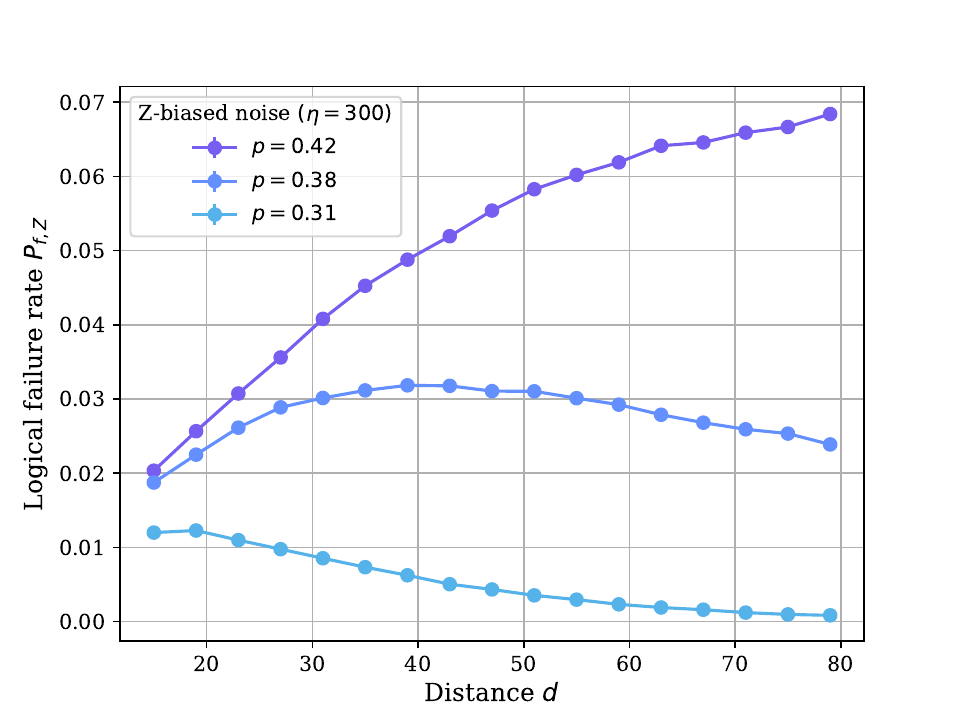}}
\caption{Estimated threshold values versus code-distance $d$, for the XZZX code, as in Fig. \ref{fig:thresholds}, at bias $\eta=300$, showing non-convergence of thresholds due to $d\ll\eta$ (a). Constant error rate cuts for failure rates versus code-distance, showing non-monotonous behavior due to finite size corrections. Each data-point averaged over 600,000 syndromes (b, c).}
\label{fig:eta300}
\end{figure*}

This appendix complements the results shown in Fig.~\ref{fig:thresholds}. First, Fig.\ \ref{fig:eta100} and Fig.\ \ref{fig:eta300} gives the numerical thresholds versus code distance for $\eta=100$ and $\eta=300$. Whereas for the former the different Pauli failure rates are quite well converged, for the latter they are not. To further exemplify the difficulties with extracting correct thresholds for large bias, we also show in these figures several plots of the failure rate at fixed error rates, as a function of code distance. For, example, in Fig.\ \ref{fig:eta300}b, for $p=0.46$, up to code distance $d\lesssim 50 $ the total failure rate decreases with code distance, suggesting that the threshold is above $p=0.46$. For larger code distances, the increasing failure rates instead suggest that the threshold is below $p=0.46$. For the logical bit-flip failure rate, Fig.\ \ref{fig:eta300}c, the opposite trend is observed, where $p=0.38$ appears to be above the threshold up to code distances $d\lesssim 40 $. 


\begin{figure*}\centering
\vspace{-0.3cm}
\subfloat[]{\includegraphics[width=0.39\linewidth, trim={0.2cm 0.15cm 1.6cm 0.4cm},clip]{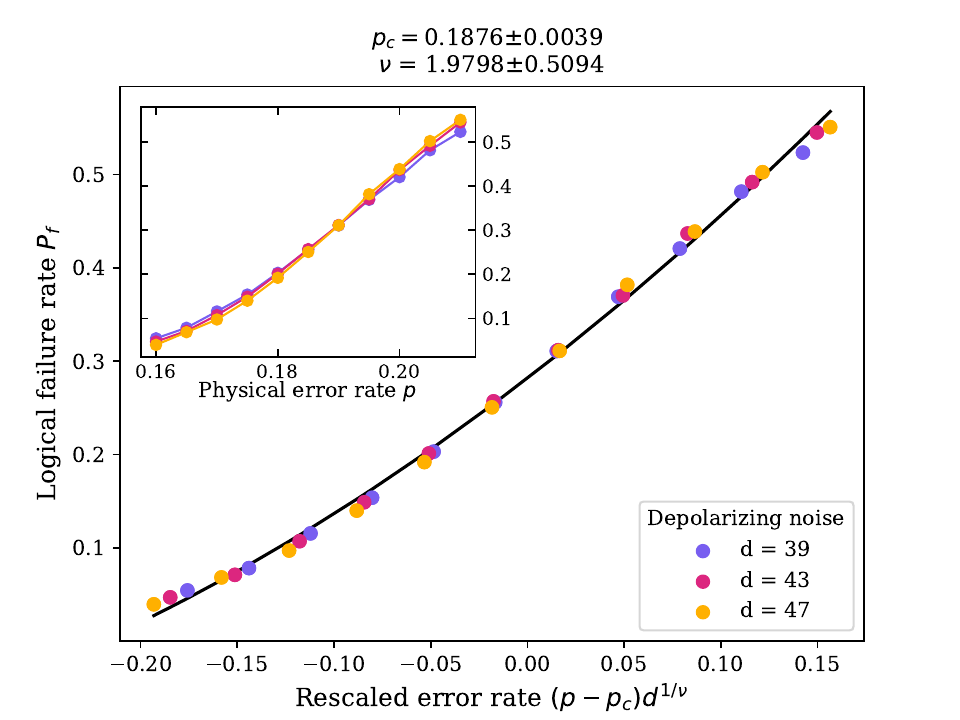}}\hspace{2cm}
\subfloat[]{\includegraphics[width=0.39\linewidth, trim={0.2cm 0.15cm 1.6cm 0.4cm},clip]{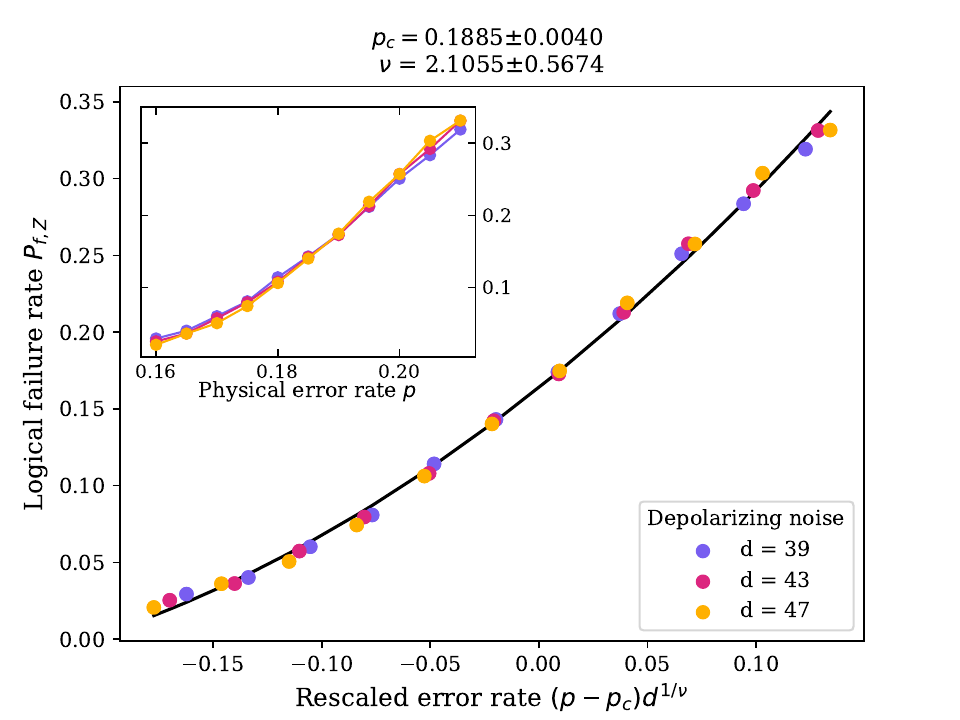}}\par
\vspace{-0.3cm}
\subfloat[]{\includegraphics[width=0.39\linewidth, trim={0.2cm 0.15cm 1.6cm 0.4cm},clip]{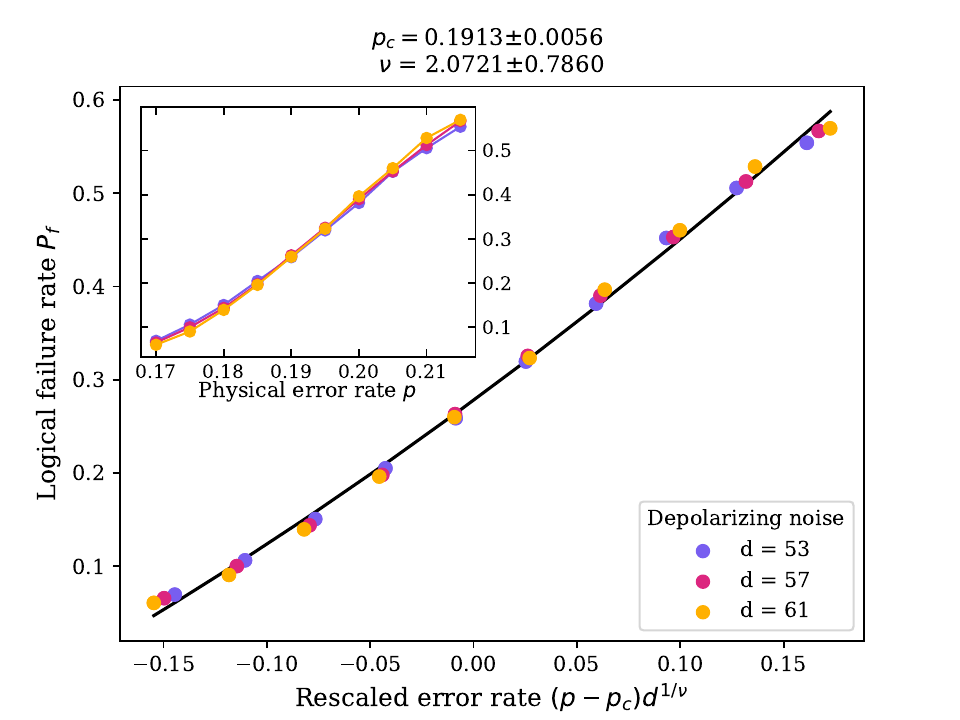}}\hspace{2cm}
\subfloat[]{\includegraphics[width=0.39\linewidth, trim={0.2cm 0.15cm 1.6cm 0.4cm},clip]{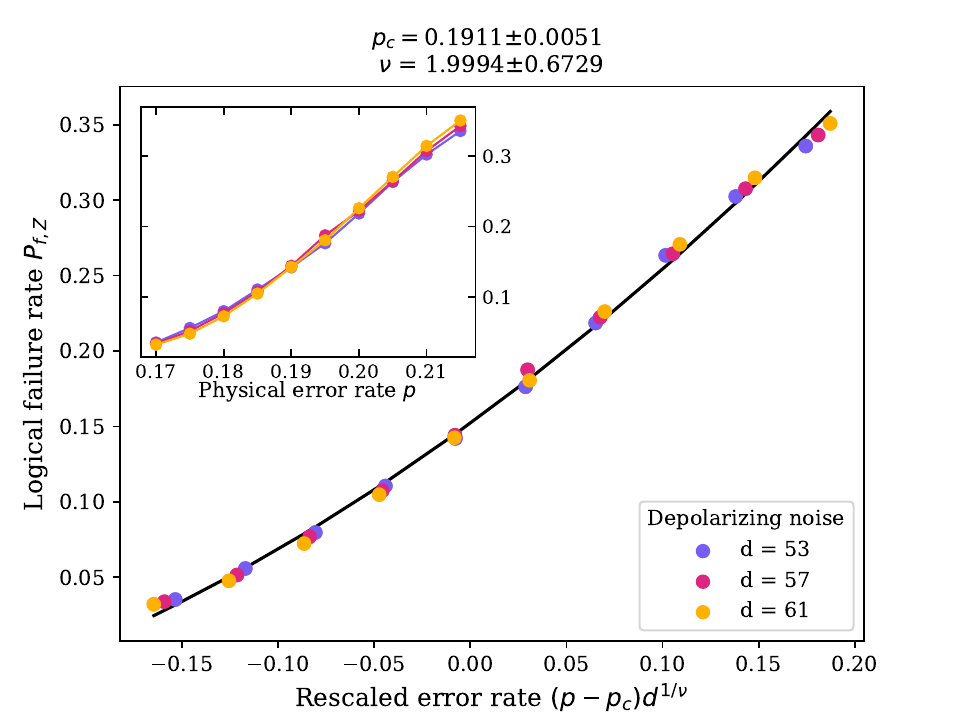}}\par
\vspace{-0.3cm}
\subfloat[]{\includegraphics[width=0.39\linewidth, trim={0.2cm 0.15cm 1.6cm 0.4cm},clip]{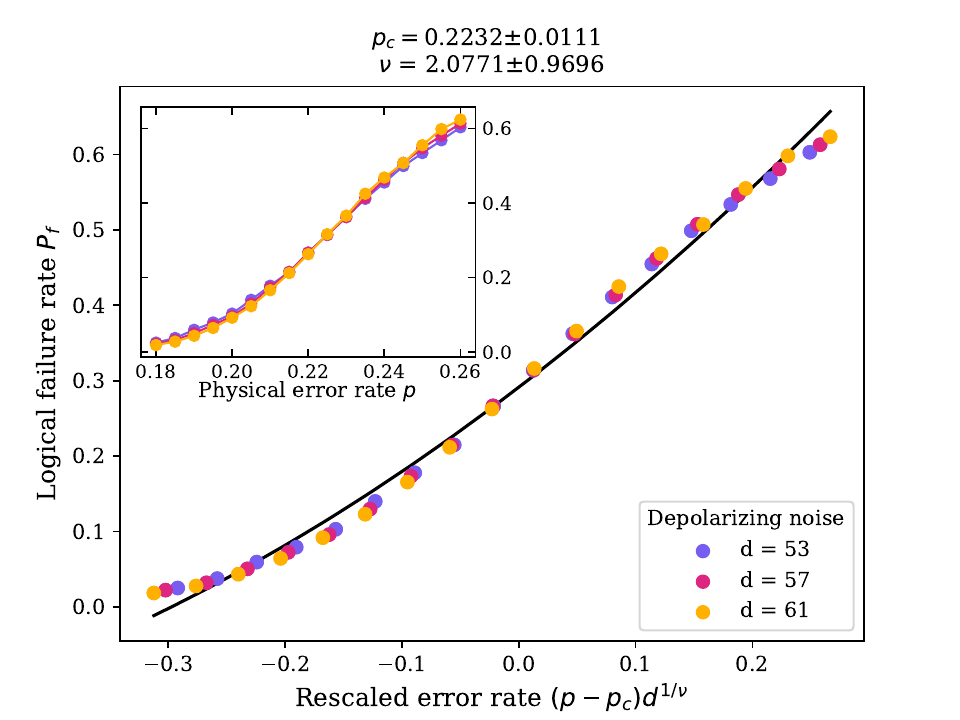}}\hspace{2cm}
\subfloat[]{\includegraphics[width=0.39\linewidth, trim={0.2cm 0.15cm 1.6cm 0.4cm},clip]{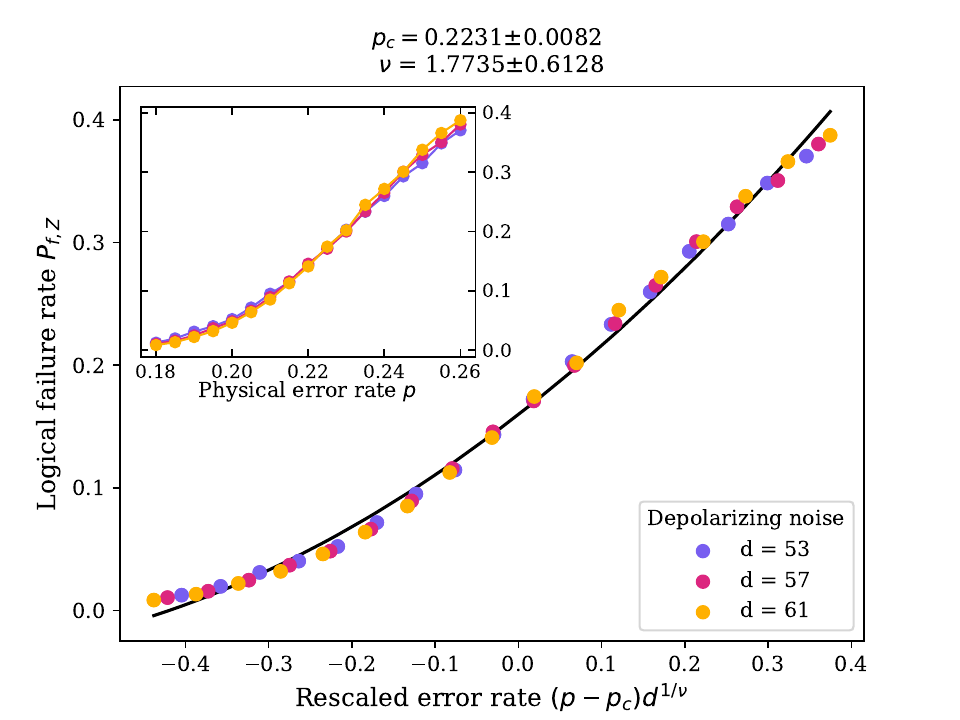}}\par
\vspace{-0.3cm}
\subfloat[]{\includegraphics[width=0.39\linewidth, trim={0.2cm 0.15cm 1.6cm 0.4cm},clip]{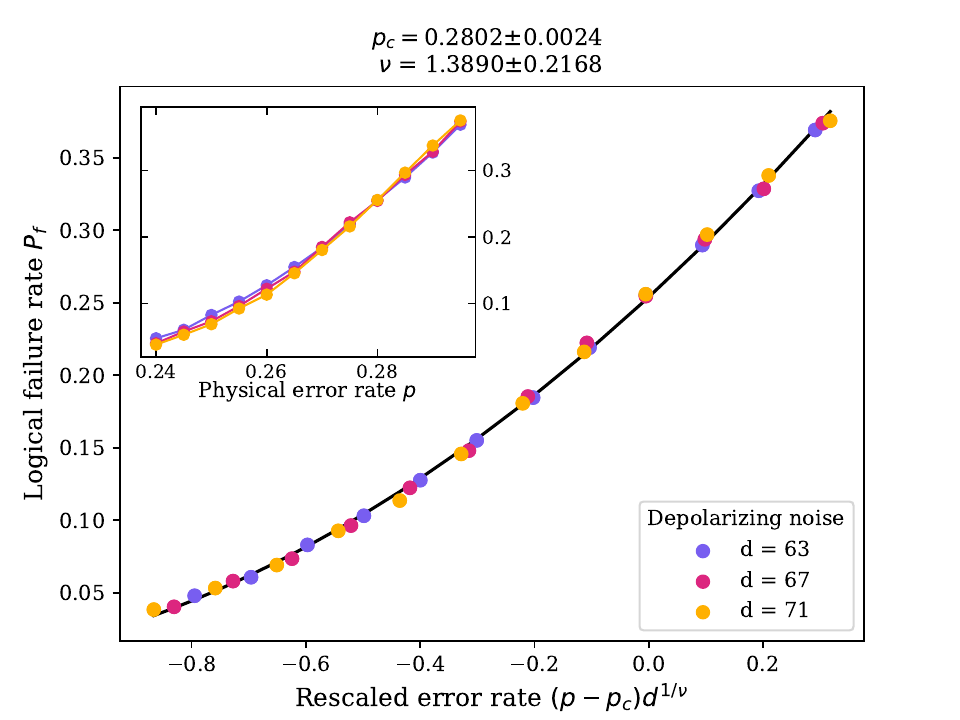}}\hspace{2cm}
\subfloat[]{\includegraphics[width=0.39\linewidth, trim={0.2cm 0.15cm 1.6cm 0.4cm},clip]{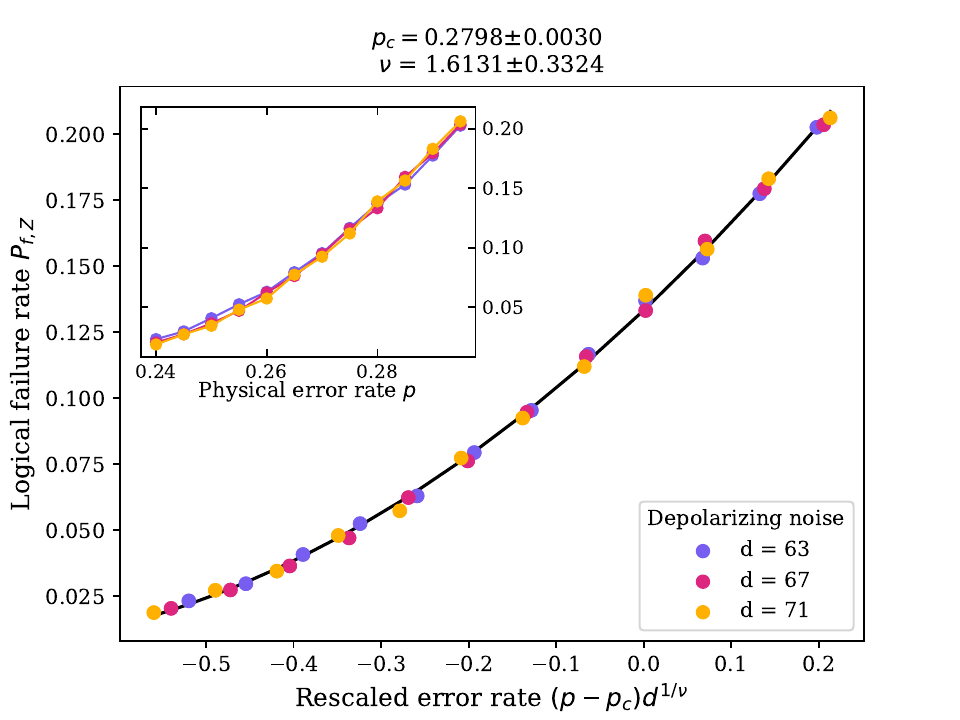}}
\vspace{-0.3cm}
\caption{Fits of the raw data (inset) to the scaling form, Eqn.\ \ref{eq:fit2}, for $\eta=0.5$ (a, b), $\eta=1$ (c, d), $\eta=3$ (e, f), $\eta=10$ (g, h), providing the threshold values shown in Fig.\ \ref{fig:thresholds}.  Similar fits (not shown) give threshold values for $P_{fX}$ and $P_{fY}$.}
\label{fig:tfdepto10}
\end{figure*}

\begin{figure*}\centering
\subfloat[]{\includegraphics[width=.49\linewidth, trim={0.2cm 0cm 1.6cm 0.4cm},clip]{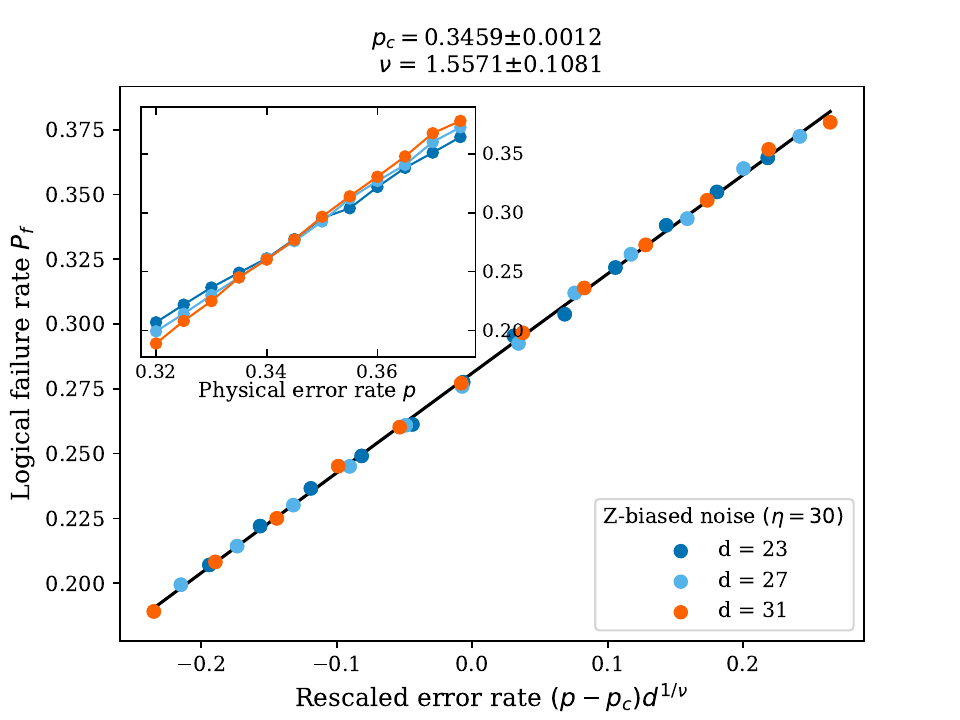}}\hfill
\subfloat[]{\includegraphics[width=.49\linewidth, trim={0.2cm 0cm 1.6cm 0.4cm},clip]{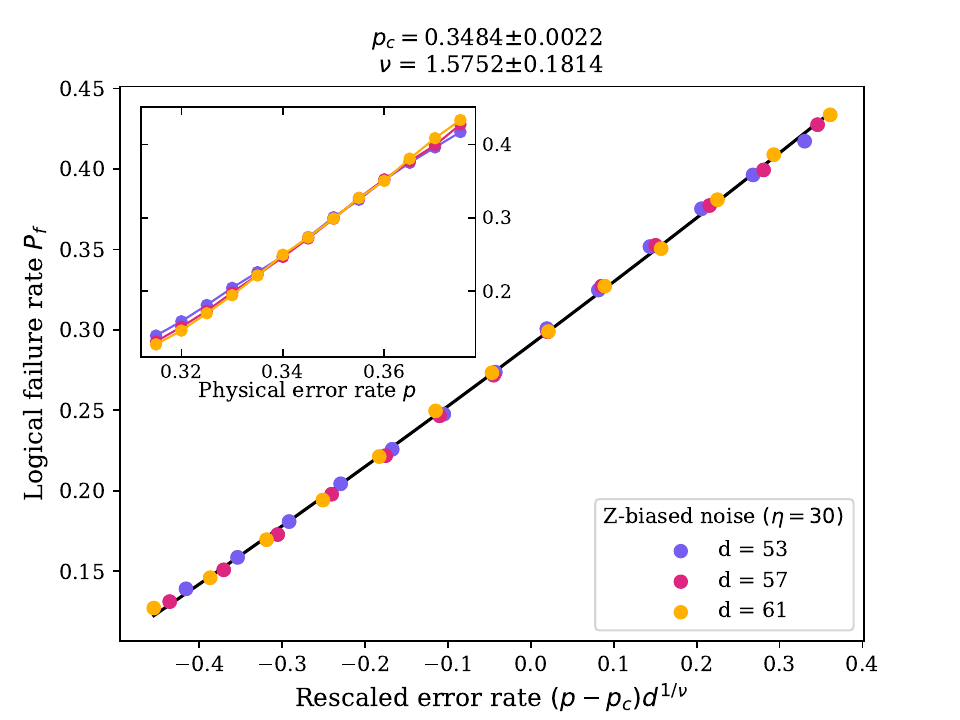}}\par
\subfloat[]{\includegraphics[width=.49\linewidth, trim={0.2cm 0cm 1.6cm 0.4cm},clip]{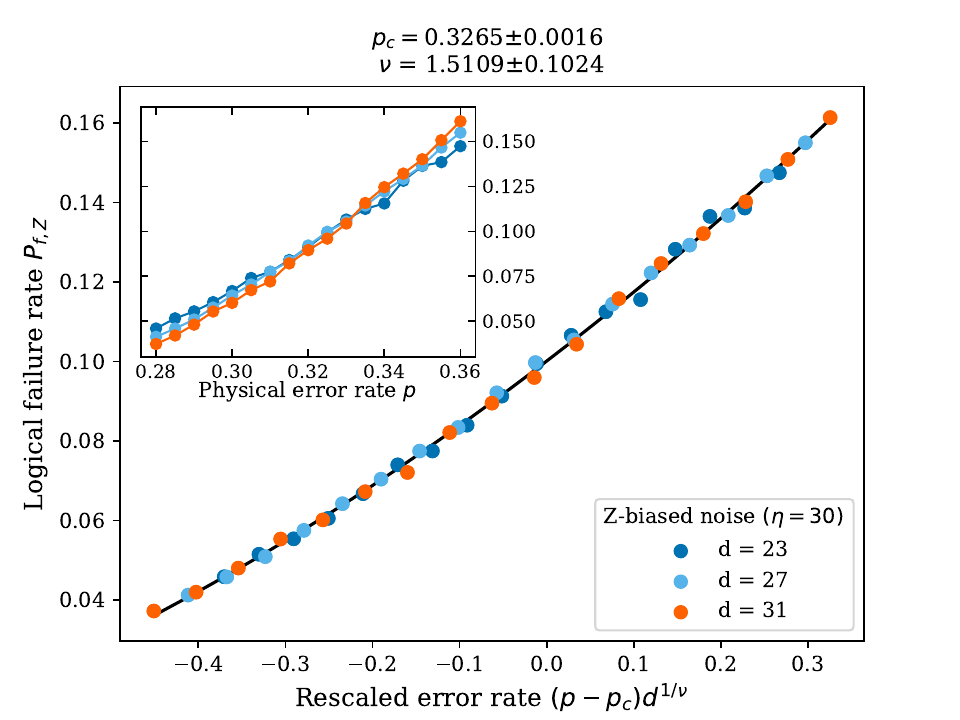}}\hfill
\subfloat[]{\includegraphics[width=.49\linewidth, trim={0.2cm 0cm 1.6cm 0.4cm},clip]{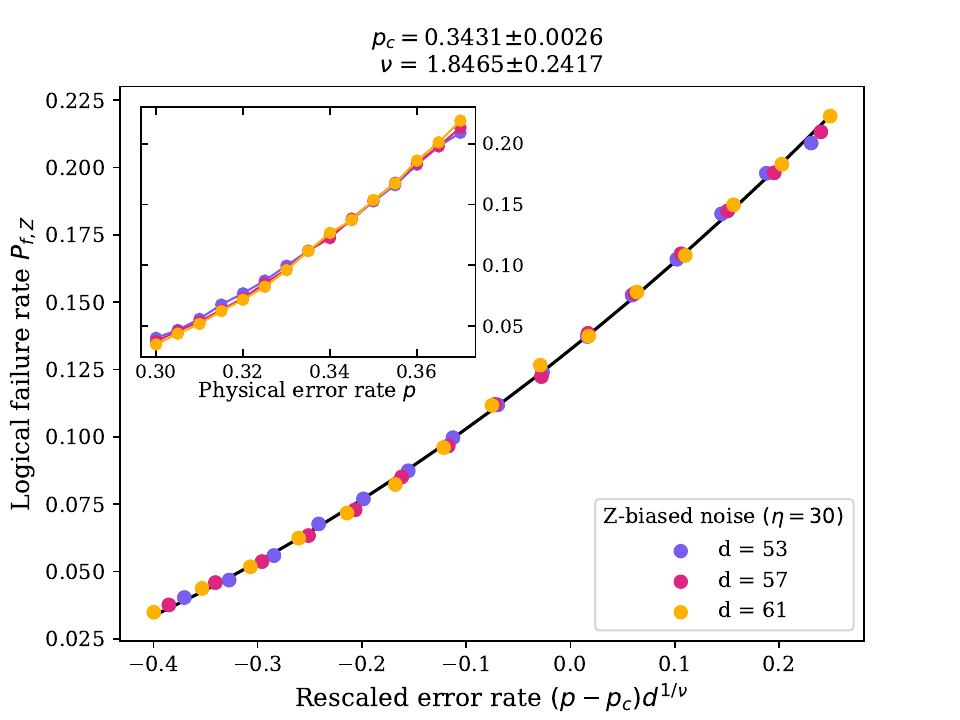}}
\caption{Fits of the raw data (inset) to the scaling form, Eqn.\ \ref{eq:fit2}, for $\eta=30$, for two different sets of code-distances and for the total, (a, b) and bit-flip, (c, d) failure rates, providing data for Fig.\ \ref{fig:thresholds}a.  Similar fits (not shown) give threshold values for $P_{fX}$ and $P_{fY}$.}
\label{fig:tf30}
\end{figure*}

\begin{figure*}\centering
\subfloat[]{\includegraphics[width=.49\linewidth, trim={0.2cm 0cm 1.6cm 0.4cm},clip]{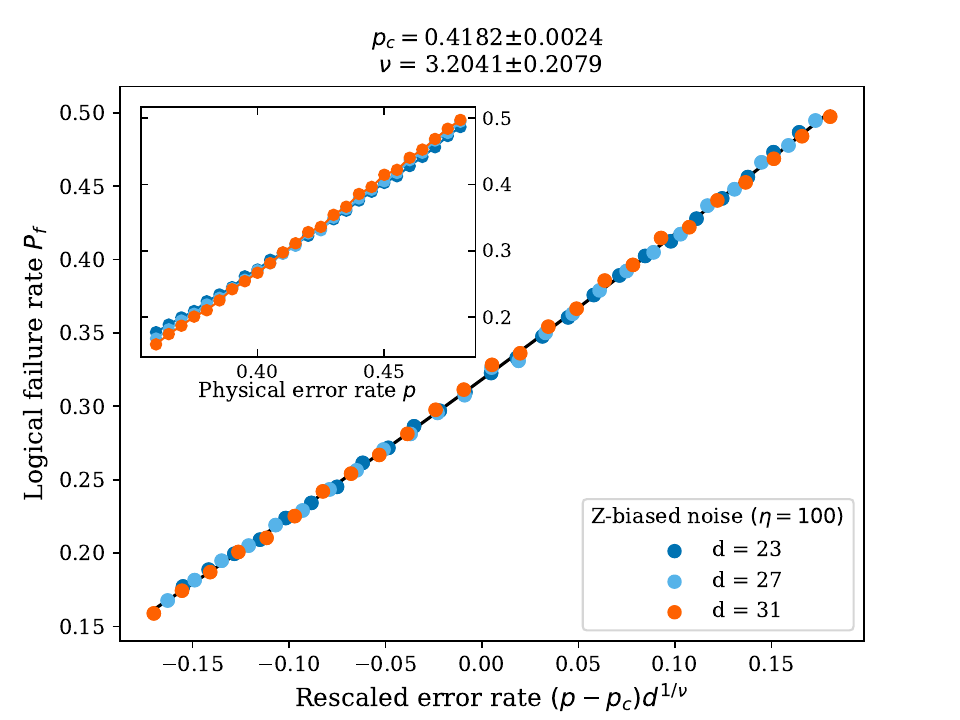}}\hfill
\subfloat[]{\includegraphics[width=.49\linewidth, trim={0.2cm 0cm 1.6cm 0.4cm},clip]{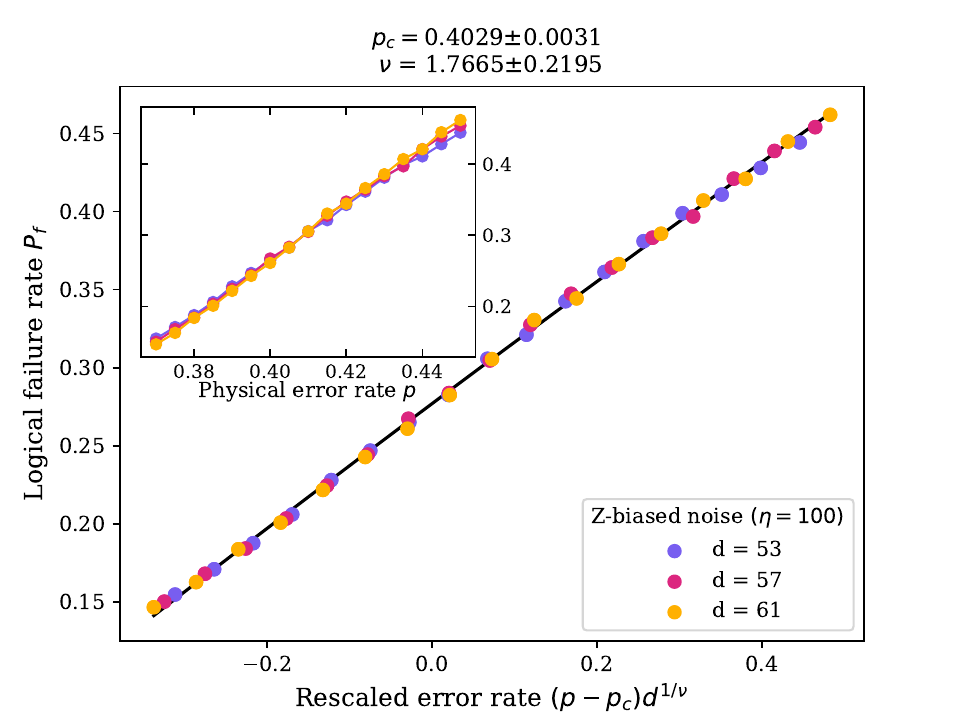}}\par
\subfloat[]{\includegraphics[width=.49\linewidth, trim={0.2cm 0cm 1.6cm 0.4cm},clip]{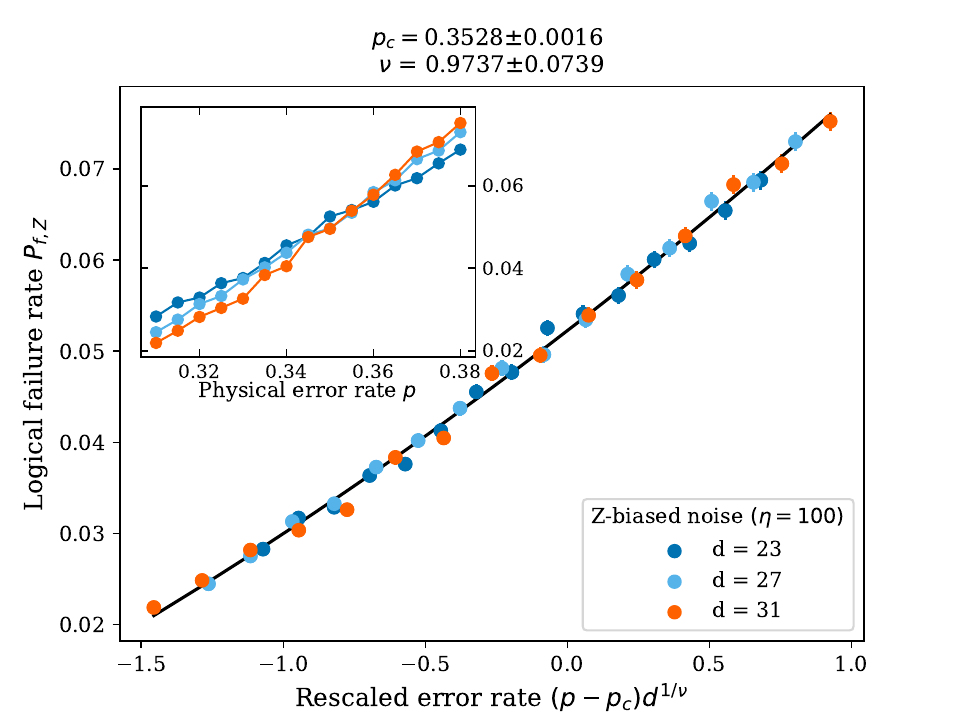}}\hfill
\subfloat[]{\includegraphics[width=.49\linewidth, trim={0.2cm 0cm 1.6cm 0.4cm},clip]{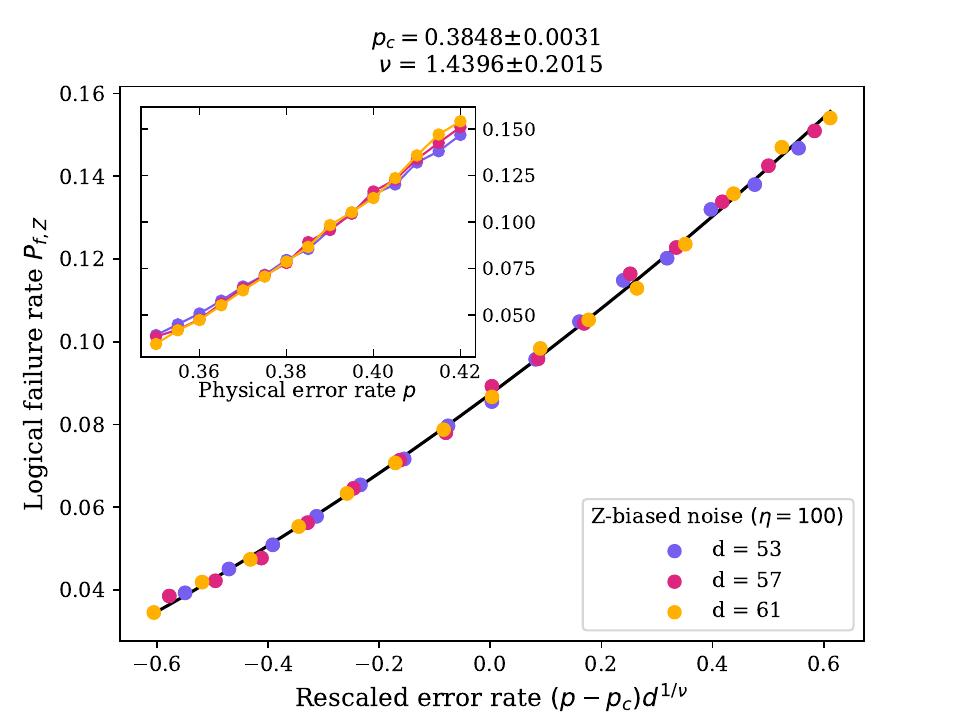}}
\caption{Fits of the raw data (inset) to the scaling form, Eqn.\ \ref{eq:fit2}, for $\eta=100$, for two different sets of code-distances and for  the total, (a, b) and bit-flip, (c, d)  failure rates, providing data for Fig.\ \ref{fig:eta100}a.  Similar fits (not shown) give threshold values for $P_{fX}$ and $P_{fY}$.}
\label{fig:tf100}
\end{figure*}

\begin{figure*}\centering
\subfloat[]{\includegraphics[width=.49\linewidth, trim={0.2cm 0cm 1.6cm 0.4cm},clip]{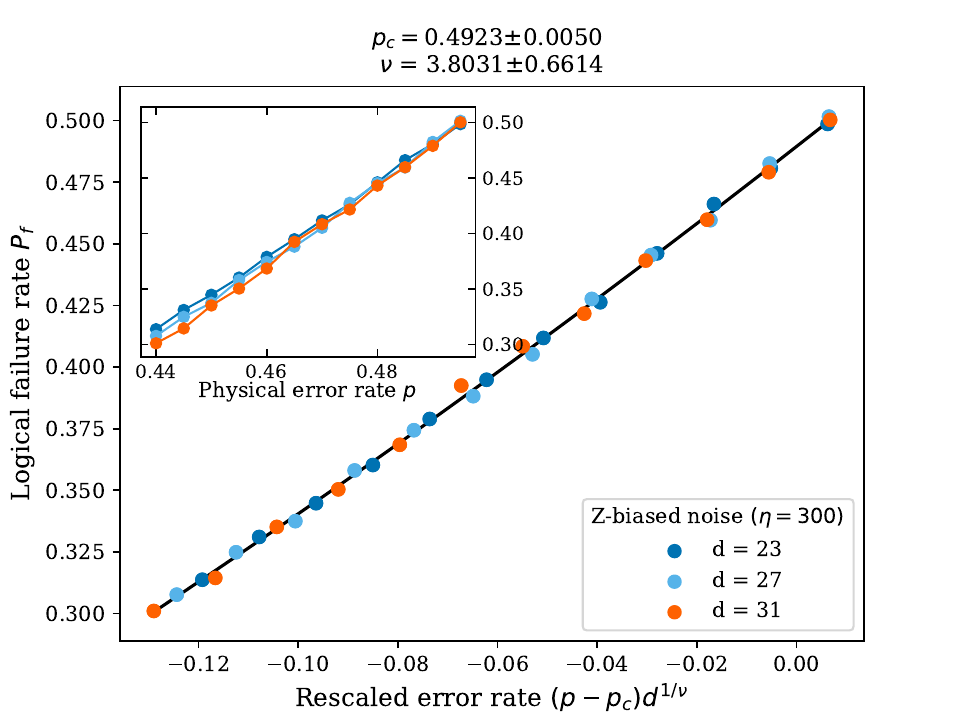}}\hfill
\subfloat[]{\includegraphics[width=.49\linewidth, trim={0.2cm 0cm 1.6cm 0.4cm},clip]{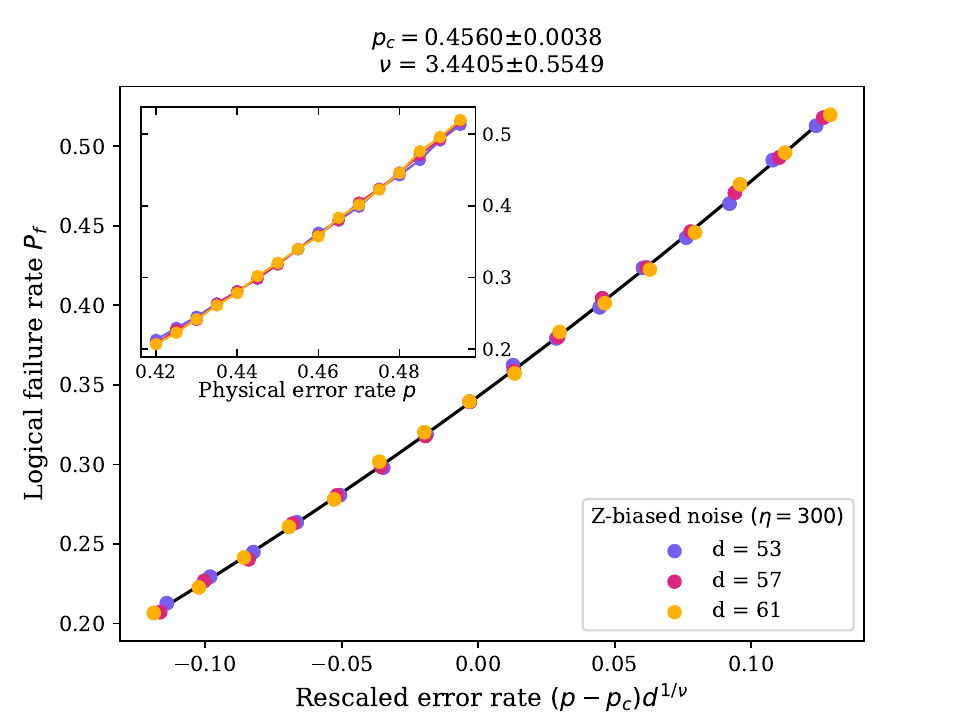}}\par
\subfloat[]{\includegraphics[width=.49\linewidth, trim={0.2cm 0cm 1.6cm 0.4cm},clip]{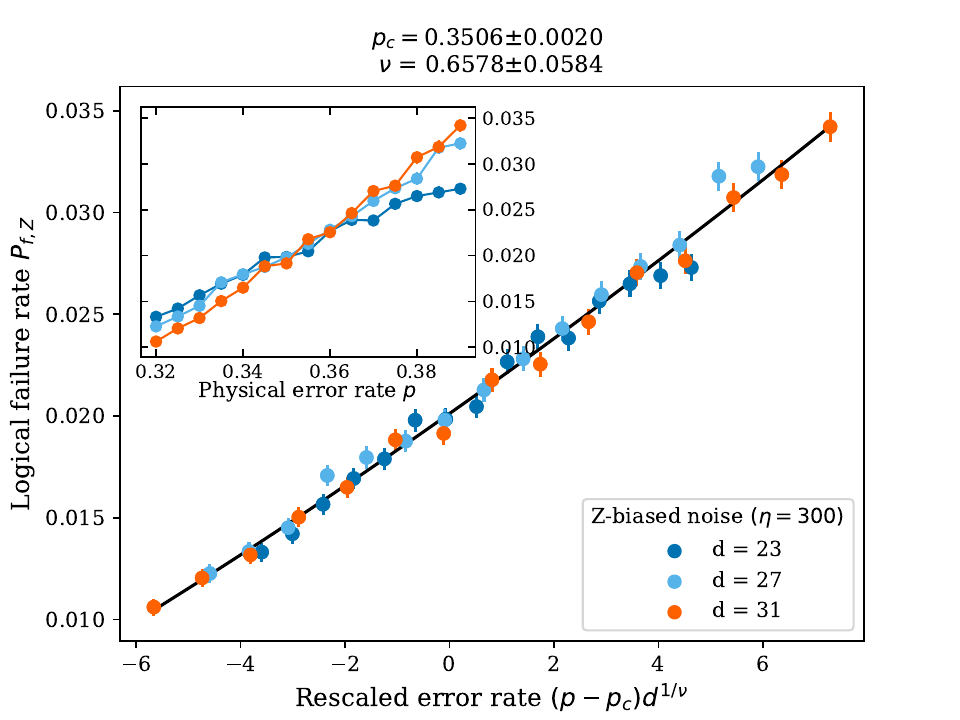}}\hfill
\subfloat[]{\includegraphics[width=.49\linewidth, trim={0.2cm 0cm 1.6cm 0.4cm},clip]{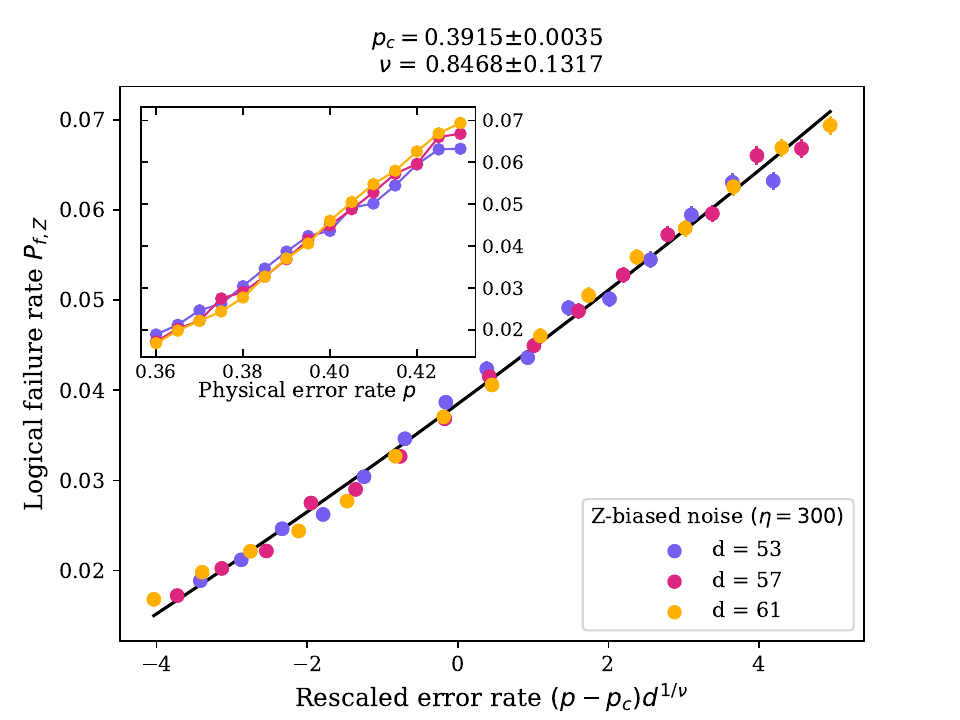}}
\caption{Fits of the raw data (inset) to the scaling form, Eqn.\ \ref{eq:fit2}, for $\eta=300$, for two different sets of code-distances and for the the total, (a, b) and bit-flip, (c, d)  failure rates, providing data for Fig.\ \ref{fig:eta300}a.  Similar fits (not shown) give threshold values for $P_{fX}$ and $P_{fY}$.}
\label{fig:tf300}
\end{figure*}

\begin{figure*}\centering
\subfloat[]{\includegraphics[width=.49\linewidth, trim={0.2cm 0cm 1.6cm 0.4cm},clip]{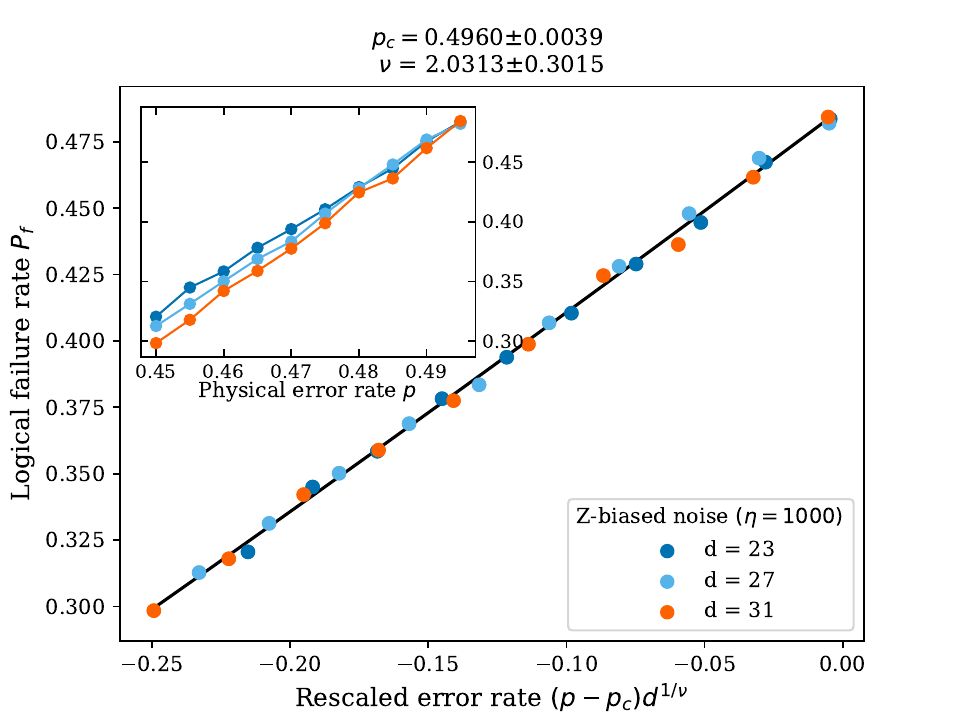}}\hfill
\subfloat[]{\includegraphics[width=.49\linewidth, trim={0.2cm 0cm 1.6cm 0.4cm},clip]{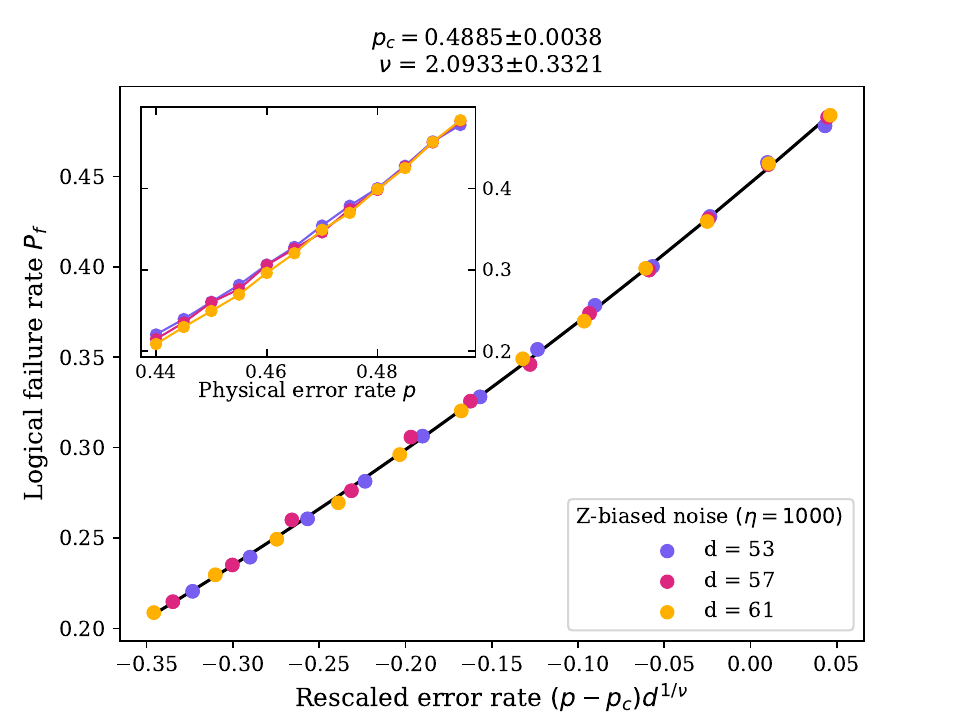}}\par
\subfloat[]{\includegraphics[width=.49\linewidth, trim={0.2cm 0cm 1.6cm 0.4cm},clip]{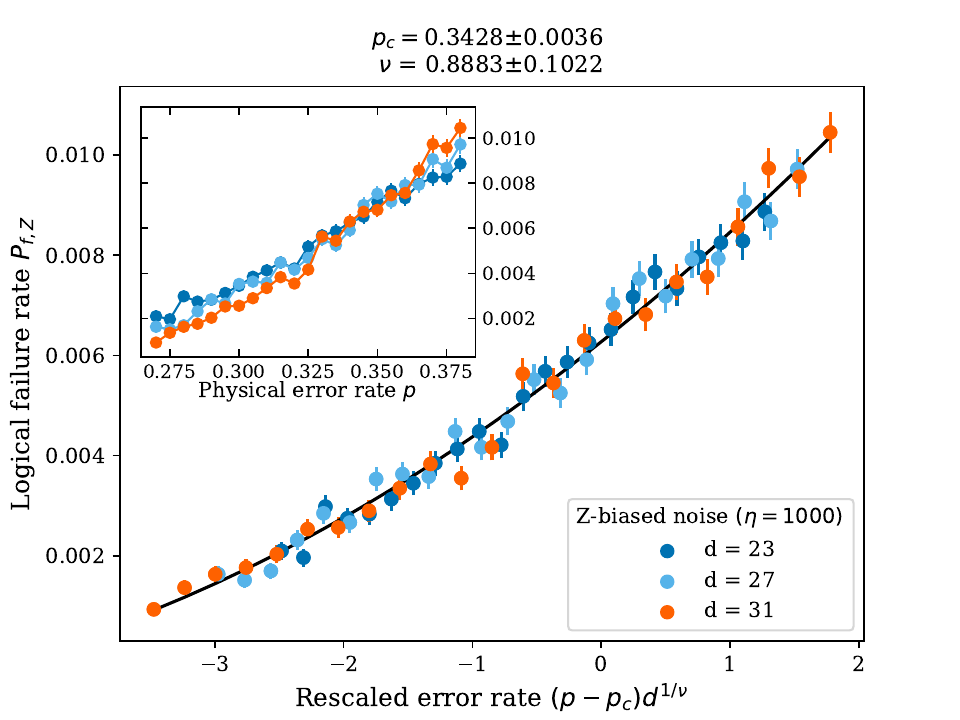}}\hfill
\subfloat[]{\includegraphics[width=.49\linewidth, trim={0.2cm 0cm 1.6cm 0.4cm},clip]{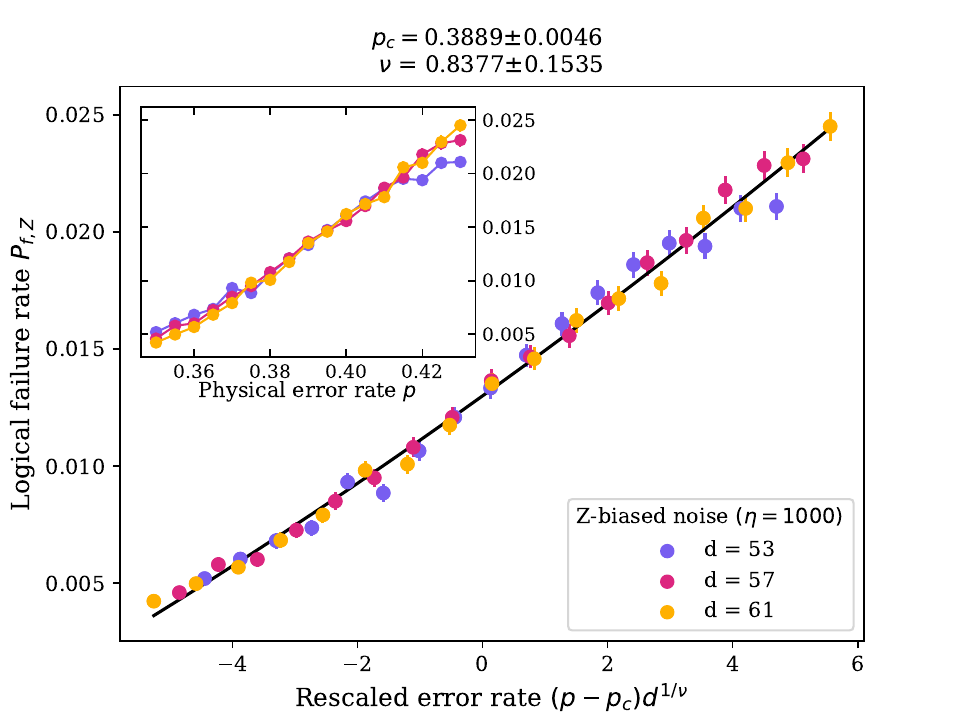}}
\caption{Fits of the raw data (inset) to the scaling form, Eqn.\ \ref{eq:fit2}, for $\eta=1000$, for two different sets of code-distances and for the the total, (a, b) and bit-flip, (c, d)  failure rates, providing data for Fig.\ \ref{fig:thresholds}b.  Similar fits (not shown) give threshold values for $P_{fX}$ and $P_{fY}$.}
\label{fig:tf1000}
\end{figure*}

The remaining figures, Fig.\ \ref{fig:tfdepto10}-\ref{fig:tf1000}, give examples of the threshold fits. As described in the main text, thresholds are extracted by using the finite size scaling fit on a set of similar code distances. Each set of code-distances and logical Pauli rates, as well as the total failure rate, are fit to the form (Eq.\ \ref{eq:fit2}) $P_f=A+Bx+Cx^2$, with $x=(p-p_c)d^{1/\nu}$, with five fitting parameters $p_c,\nu,A,B$, and $C$. For the simulation for each error $\eta$ and each set of code-distances $[d-4,d,d+4]$, different ranges of error rates that capture the threshold are used, but with a fixed interval $\Delta p = 0.005$; for every error rate, 60000 random syndromes are decoded using bond dimension $\chi=8$ for $\eta \geq 30$ and 30000 random syndromes using bond dimension $\chi=16$ for $\eta<30$. (These parameters are consistent with those used in \cite{Ataides2021XZZX}.) The plots illustrate (see, e.g.\ Fig. \ref{fig:tf300}a-b and c-d) the substantial drift for large bias of the crossing point between the failure rates with code distance and the corresponding best value for the threshold value $p_{c}$. In addition, both the value of $p_{c}$ and the exponent $\nu$, extracted from the total failure rate $P_f$ or from the bit-flip failure rate $P_{fZ}$ should coincide, given a single phase transition, which is clearly not the case for code distances $d\ll \eta$.

\end{document}